# Observations of ionized carbon towards the gamma-ray-bright supernova remnant RX J1713.7−3946

A. R. Thakur [1★] G. P. Rowell,[1] R. Higgins,[2] M. Burton [3,4] Y. Fukui,[5] S. Einecke [1] H. Sano [6] G. Wong,[4,7] M. D. Filipović[7] and C. Braiding[4]

[1]*School of Physics, Chemistry and Earth Sciences, Adelaide University, North Terrace, Adelaide SA 5005, Australia*
[2]*Department of Physics, University of Cologne, Albertus-Magnus-Platz, D-50923 Köln, Germany*
[3]*Armagh Observatory and Planetarium, College Hill, Armagh BT61 9DG, UK*
[4]*School of Physics, University of New South Wales, Sydney NSW 2052, Australia*
[5]*Department of Physics (Graduate School of Science), Nagoya University, Nagoya, Furocho, Chikusa Ward, Nagoya, Aichi 464-8601, Japan*
[6]*Applied Mathematics and Physics Division, Gifu University, 1-1 Yanagido, Gifu 501-1193, Japan*
[7]*School of Science, Western Sydney University, Penrith NSW 2751, Australia*

## ABSTRACT

The origin of Galactic cosmic rays (CRs), particularly at sub-GeV energies, remains uncertain due to their difficulty in being traced. One possible tracer of low-energy CRs is ionized carbon ([C II]), which may be enhanced in regions affected by CR-induced ionization. We present observations of [C II] 158 μm line emission across the $\gamma$-ray-bright supernova remnant (SNR) RX J1713.7−3946 obtained with SOFIA. These data are compared with molecular and atomic gas from Mopra ($^{12}$CO), Nanten ($^{12}$CO) and SGPS (H I), and high-energy emission from TeV $\gamma$-ray H.E.S.S. and X-ray *XMM–Newton*. We find [C II] emission follows the atomic gas more closely than molecular gas, particularly in regions near the SNR shock front. Ratios of [C II] intensity to gas column density vary across the remnant, with peak values coinciding with regions of enhanced TeV $\gamma$-ray emission. To put the [C II] in a wider context, we examined pointings from the *Herschel* GOT C+ survey. Our analysis of the GOT C+ data shows no significant difference in the I[C II]/I[$^{12}$CO] ratio for observations towards star formation regions, H II regions and SNRs. To assess the origin of carbon ionization, we model [C II] emissivity using the photoionization code CLOUDY. We find UV photons and CRs can produce comparable levels of [C II] emission, assuming low-energy CRs ($< 1$ GeV) are accelerated and trapped within the SNR shock. These results highlight the potential of [C II] as a tracer of CR ionization and the importance of combining multiwavelength observations to probe CR interactions in SNRs.

## 1 INTRODUCTION

The origin of Galactic cosmic rays (CRs) has been debated for decades (V. L. Ginzburg & S. I. Syrovatsky 1965; S. Gabici et al. 2019). At energies $\lesssim 10^{19}$ eV, it is impossible to directly trace these particles to their origin as their trajectories are affected by magnetic fields in our Galaxy. CRs with energies $\geq 10$ GeV may be traced by following $>0.1$ GeV $\gamma$-ray emission observed with telescopes like the High Energy Stereoscopic System (H.E.S.S.), MAGIC (C. Bigongiari 2005), the Large High Altitude Air Shower Observatory (Z. Cao et al. 2019), the Very Energetic Radiation Imaging Telescope Array System (J. Holder et al. 2006) and the *Fermi* Large Area Telescope (*Fermi*-LAT; W. B. Atwood et al. 2009). These $\gamma$-rays are produced by interaction of multi-GeV CRs with gas in the interstellar medium.

The sub-GeV $\gamma$-ray emission from the lowest-energy Galactic CRs ($< 10$ GeV) corresponds to the poorest angular resolution of *Fermi*-LAT ($> 1^{\circ}$; W. B. Atwood et al. 2009). While TeV $\gamma$-ray facilities like H.E.S.S. have sufficient angular resolution to pinpoint CR origins, the GeV $\gamma$-ray resolution of *Fermi*-LAT does not. Therefore, it is not possible to use sub-GeV $\gamma$-ray emission to accurately trace the origins of low-energy sub-GeV CRs. However, these CRs can ionize gas clouds, offering a potential way to trace them. Table A1 summarizes the energetic photons and CR populations relevant to this work. Far-UV (FUV) photons and X-rays ionize carbon directly but their influence can be limited by extinction and column density, respectively. While CRs span a broad energy range, multi-GeV CRs are traced indirectly through the $\gamma$-ray emission they produce, whereas the low-energy ($\lesssim 1$ GeV) CRs of interest here are not easily traced by $\gamma$-rays. In the MeV band, COMPTEL measured the diffuse Galactic continuum (0.75−30 MeV; A. W. Strong et al. 1994), but this emission is a blend of non-thermal bremsstrahlung from low-energy CR electrons, inverse Compton scattering and positronium continuum, making it difficult to isolate the low-energy CR proton component that drives ionization. This, combined with the poor angular resolution of MeV instruments ($\sim 1^{\circ}$ for COMPTEL; A. W. Strong et al. 1994), prevents the low-energy CRs from being localized to specific clouds or remnants, motivating the use of [C II] as a complementary, spatially resolved tracer of their ionizing effect. We note that the most widely used tracers of ionization are $DCO^+$ and $HCO^+$ (M. Guelin et al. 1977). However, $DCO^+$ only represents a small fraction of the ionized gas within gas clouds, as it can only be detected in cores of dense gas clouds (E. Bron et al. 2021).

★ Email: adnaan@thakur.com

For many decades, supernova remnants (SNRs) have been suggested as a source of Galactic CRs (V. L. Ginzburg & S. I. Syrovatsky 1965). SNR shocks are the leading candidate sites for diffusive shock acceleration of particles. The kinetic energy released per supernova ($\sim 10^{51}$ erg), combined with the Galactic supernova rate (once every 40 yr; W. Baade & F. Zwicky 1934; V. L. Ginzburg & S. I. Syrovatsky 1965; G. A. Tammann, W. Loeffler & A. Schroeder 1994), is sufficient to sustain the energy density of Galactic CRs. Young SNRs are particularly valuable, as freshly accelerated sub-GeV CRs could remain confined near the shock and any adjacent dense target gas, making these remnants ideal objects to study the origin of CRs. In this paper, we will look at the use of ionized carbon ($C^+$) to potentially trace sub-GeV CRs in a young supernova remnant. $C^+$ is usually detected in the outer layers of clouds, and has been used to study interactions between gas clouds and examine star formation regions (N. Schneider et al. 2023). The spatial overlap between $C^+$ and $\gamma$-ray emission could indicate the presence of CR ionization.

E. Bayet et al. (2011) modelled the abundances of different ions as a function of the CR-induced ionization rate. They found the fractional abundance of $C^+$ increased with the CR ionization rate. B. Gaches, T. Bisbas & S. Bialy (2022) modelled molecular clouds and quantified variations in column densities and line emission of carbon cycle species using different CR ionization rates. E. Bayet et al. (2011) and B. Gaches et al. (2022) suggest $C^+$ (and its 1.9 THz fine structure line [C II]) can be used to trace the CR ionization rate in gas clouds.

Since low-energy CRs might not have sufficient energy to penetrate through to the centre of the cloud cores, they could lose their energy after ionizing the gas in the outer layers (A. Tielens 2005; S. Gabici 2017). As a result, we may see the strongest $C^+$ emission, characterized by the 1.9 THz [C II] fine structure line emitted from the outer layers of the gas, which may impact the ratio of [C II] to molecular and atomic gas in the interstellar medium (ISM). If the gas is primarily molecular and dense, it would be difficult for the low-energy CRs to ionize the gas.

There are other sources of ionized carbon besides that from CRs, such as FUV photons and X-rays. A. G. G. M. Tielens & D. Hollenbach (1985); M. G. Burton, D. J. Hollenbach & A. G. G. M. Tielens (1990); A. N. Heays, A. D. Bosman & E. F. Dishoeck (2017) looked at the photoionization and photodissociation of atoms and molecules. Their results suggest that in the diffuse ISM at modest extinction, FUV (6–13.6 eV) photons are the dominant source of carbon ionization. Although these photons lie below the hydrogen ionization threshold, they photoionize carbon and other heavy elements, and continue to produce $C^+$ within molecular gas, forming the photodissociation region. The depth to which FUV ionization remains effective is density-dependent, persisting to $A_V \sim 10$ in dense gas ($n \sim 10^6$ cm$^{-3}$) but becoming negligible by $A_V \sim 3$ at higher densities ($n \sim 10^7$ cm$^{-3}$), beyond which CRs are the dominant ionization source (A. G. G. M. Tielens & D. Hollenbach 1985; M. G. Burton et al. 1990). X-rays are capable of travelling deep into the gas clouds and ionizing the carbon. We would expect $C^+$ produced by X-ray ionization to be distributed evenly across clouds of varying density (A. Tielens 2005). The photoionization code CLOUDY (G. J. Ferland et al. 1998) allows us to compare the [C II] emissivity produced by UV emission to that produced solely by CRs. We present a new look at ionized carbon in SNRs that will show how $C^+$ is distributed across a remnant and could help understand whether this ionized carbon is produced via UV radiation or low-energy CRs.

The SNR we have focused on is RX J1713.7−3946 (hereafter RX J1713), a relatively young SNR that is bright in both $\gamma$-rays and X-rays (Y. Fukui et al. 2012). The H.E.S.S. Galactic Plane Survey determined the shell-like shape of the $\gamma$-ray emission of the SNR (H.E.S.S. Collaboration 2018b). The SNR is approximately 1600 yr old and the shock front from the explosion has reached a velocity of 3000 km s$^{-1}$ (Y. Fukui et al. 2012). The H.E.S.S. Galactic Plane Survey identifies the location and size of RX J1713 in the Galactic plane (H.E.S.S. Collaboration 2018a). It is clear that this remnant is one of the brightest in the galaxy and as such it is believed to be a candidate to determine the origin of CRs. The age of the SNR may suggest that many of the CRs (sub-GeV energies) produced in the initial explosion may still be trapped within the SNR bubble, being accelerated via diffusive shock acceleration. Clouds within the SNR could be bombarded by CRs, leading to a high ionization rate. The ionization within clouds would lead to a potential excess of ionized carbon.

In order to search for ionized carbon across RX J1713, we used the [C II] 1.9 THz transition line emission data taken from the Stratospheric Observatory for Infrared Astronomy (SOFIA) telescope (A. Krabbe 2000). The ISM gas surrounding the remnant has been studied using the SOFIA, Mopra (M. Burton et al. 2013; C. Braiding et al. 2018; K. Cubuk et al. 2023) and Nanten (Y. Moriguchi et al. 2005; Y. Fukui 2008; Y. Fukui et al. 2021) telescopes. The study of $^{12}$CO(J = 1–0) gas, hereafter referred as $^{12}$CO, in the direction of the SNR showed that the molecular gas is in a local standard of rest velocity ($v_{LSR}$) range of $v_{LSR} = -20$ to 0 km s$^{-1}$ (Y. Fukui et al. 2012). We will use this velocity range as a guide to study the data collected by SOFIA. The H.E.S.S. gamma-ray and the *XMM–Newton* X-ray emission maps were used to determine the locations of the shock front across the SNR.

The regions from which SOFIA collected data were chosen based on the ($> 0.2$ TeV) $\gamma$-ray emission from H.E.S.S., shown in Fig. 1. While these are not the $\gamma$-rays associated with the sub-GeV CRs we are investigating, they can be used to identify regions where there are likely to be a significant number of CRs.

The maps shown in Fig. 1 are the $^{12}$CO and X-ray data, from Mopra and *XMM–Newton,* respectively. The overlaid regions are those taken from the SOFIA data collection (discussed shortly). The X-ray map shows the positions of the shock front from the initial SN explosion. Regions M1 and M3 overlap with those shock fronts. The $^{12}$CO map shows the positions of the gas clouds across the SNR.

## 2 OBSERVATIONS TOWARDS SNR RX J1713

### 2.1 SOFIA [C II]

The SOFIA, Stratospheric Observatory for Infrared Astronomy (A. Krabbe 2000), telescope was a modified Boeing 747SP aircraft

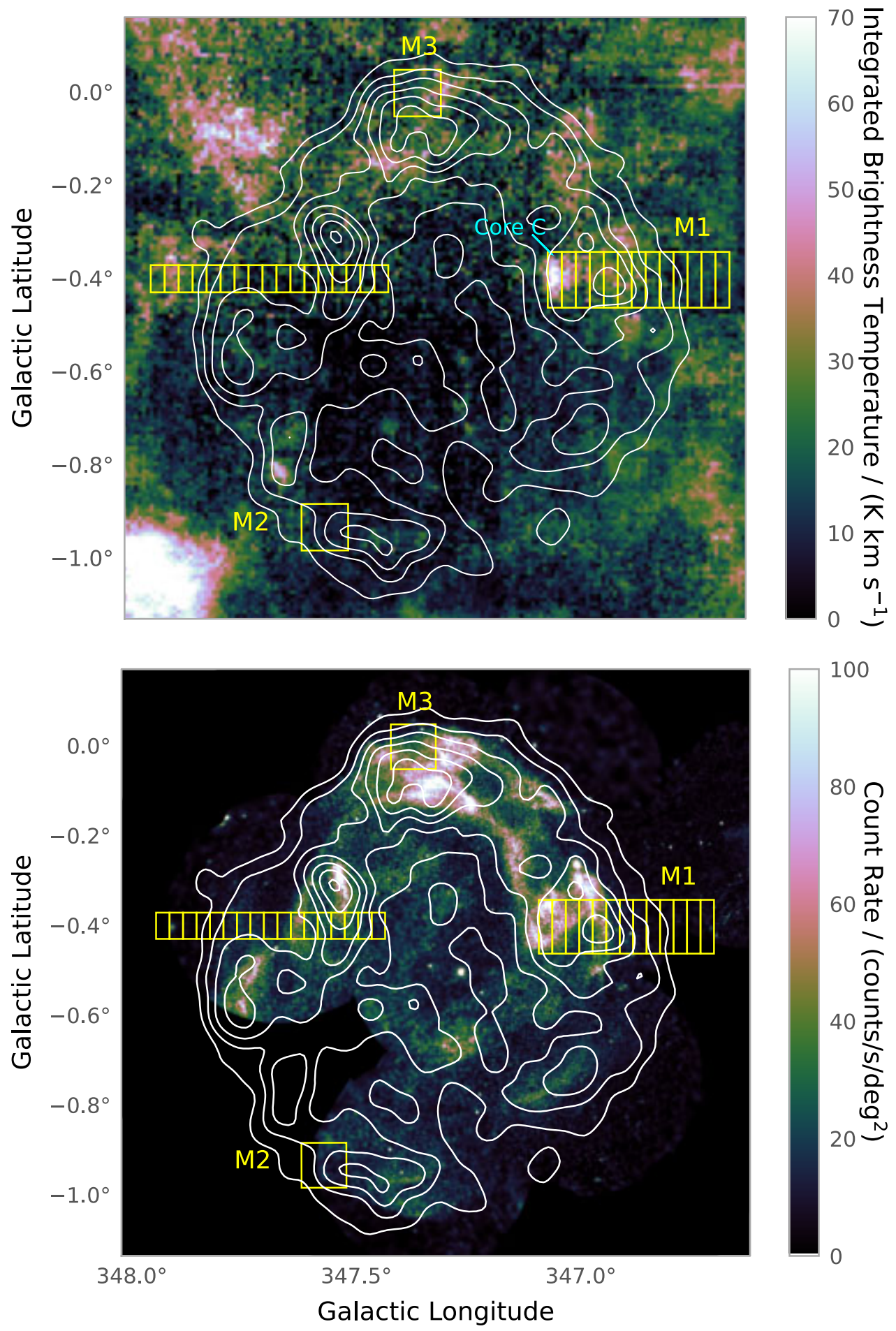

**Figure 1.** *Top panel*: Integrated Line Intensity of $^{12}$CO(1-0) transition observed by Mopra across a velocity range of $-25$ km s$^{-1}$ to 0 km s$^{-1}$. *Bottom panel*: X-ray emission observed by *XMM–Newton*, 2.0–7.2 keV. *Contours*: SNR RX J1713.7−3946 as observed in > 2 TeV H.E.S.S. $\gamma$-rays. The yellow regions show our SOFIA mapping of the [C II] spectral line. The M1 region is split into 30 sub-regions. The numbering moves from left to right, with M1-1 the first box on the left and M1-30 the last box on the right.

that houses a 2.7 m reflecting telescope with an effective diameter of 2.5 m.[1] The aircraft flew at altitudes between 11 600 and 13 700 m when the telescope collected data. At this height, the telescope is positioned above almost all of the Earth's infrared-absorbing (between 4 and 400 μm) atmosphere. As such, SOFIA was one of the best telescopes available to map the distribution of C$^+$ throughout SNR RX J1713.7−3946.

Onboard SOFIA the C$^+$ observations were undertaken with the heterodyne upGREAT receiver (C. Risacher et al. 2016). This was a dual polarization receiver in a hexagonal format, with pixels at each corner and pixel in the middle centre making 7 pixels per polarization, resulting in 14 pixels total. In addition to the C$^+$ observations, atomic oxygen (OI) was also observed in parallel however no detection was seen and the data are not presented here. The C$^+$ data were taken using the on-the-fly observation mode. The observations were taken in array mapping mode, which allowed for fast mapping of large regions (see R. Higgins et al. (2021) for more details). Data were taken over three flights as part of a southern campaign based at Christchurch, New Zealand. The RX J1713 data were observed on flight numbers 469 (2018-06-14), 472 (2018-06-19) and 474 (2018-06-20).

As the entire RX J1713 remnant was too large to map in its entirety, a sample of three regions was observed based on the available TeV $\gamma$-ray and X-ray observations. The regions (M1, M2, and M3) are shown in Fig. 1. M1 is a strip running across the SNR, M2 is a region to the south, and M3 is a region to the north of the SNR. A common OFF region was used for all observations. Given the proximity to the Galactic plane and potential for [C II] emission in the OFF position, this OFF region was checked against another OFF region further from the Galactic plane. These data were inspected during the flight, and the near OFF position was initially deemed to be free of [C II] emission. However, further analysis (discussed in Section 3) revealed evidence for [C II] emission in the OFF region. The removal of this artefact is discussed in Section 3. The location of the OFF region is shown in Table 1.

The observations covered a broad velocity range of $-150$ km s$^{-1}$ to 50 km s$^{-1}$. Since we are interested in SNR RX J1713 and the clouds adjacent to it, the primary velocity range of interest is $v_{LSR} \approx -20$ km s$^{-1}$ to 0 km s$^{-1}$ (Y. Fukui et al. 2012). The spectral resolution is $\Delta v_{LSR} \sim 0.67$ km s$^{-1}$. This resolution reveals small-scale changes in the [C II] spectra (see Figs 2 and 3).

The exposure time and region size for the SOFIA data were different for either side of the remnant. During the first flight the mapping approach was adapted to have smaller regions at high signal-to-noise ratios (note the different tile size in the M1 strip). This led to different $T_{RMS}$ values. The $T_{RMS}$ was found to be 0.65 K for the left side (longitude > 347.3°) and 0.82 K for the right side (longitude < 347.3°). While they are different, both values are low compared to the signal strength (a maximum of ∼15 K), which makes it easy to identify changes in the I[C II] signal, relative to the Mopra $^{12}$CO(J = 1–0) data.

**Table 1.** The central Galactic coordinates and $v_{LSR}$ (velocity with respect to local standard of rest) width of the observed regions of SNR RX J1713.

| Region | Galactic longitude (deg) | Galactic latitude (deg) | $v_{LSR}$ (km s$^{-1}$) |
|---|---|---|---|
| M1 | 347.3 | −0.4 | −150 to 50 |
| M2 | 347.6 | −0.9 | −150 to 50 |
| M3 | 347.4 | 0.0 | −150 to 50 |
| OFF | 347.4 | −1.2 | −150 to 50 |

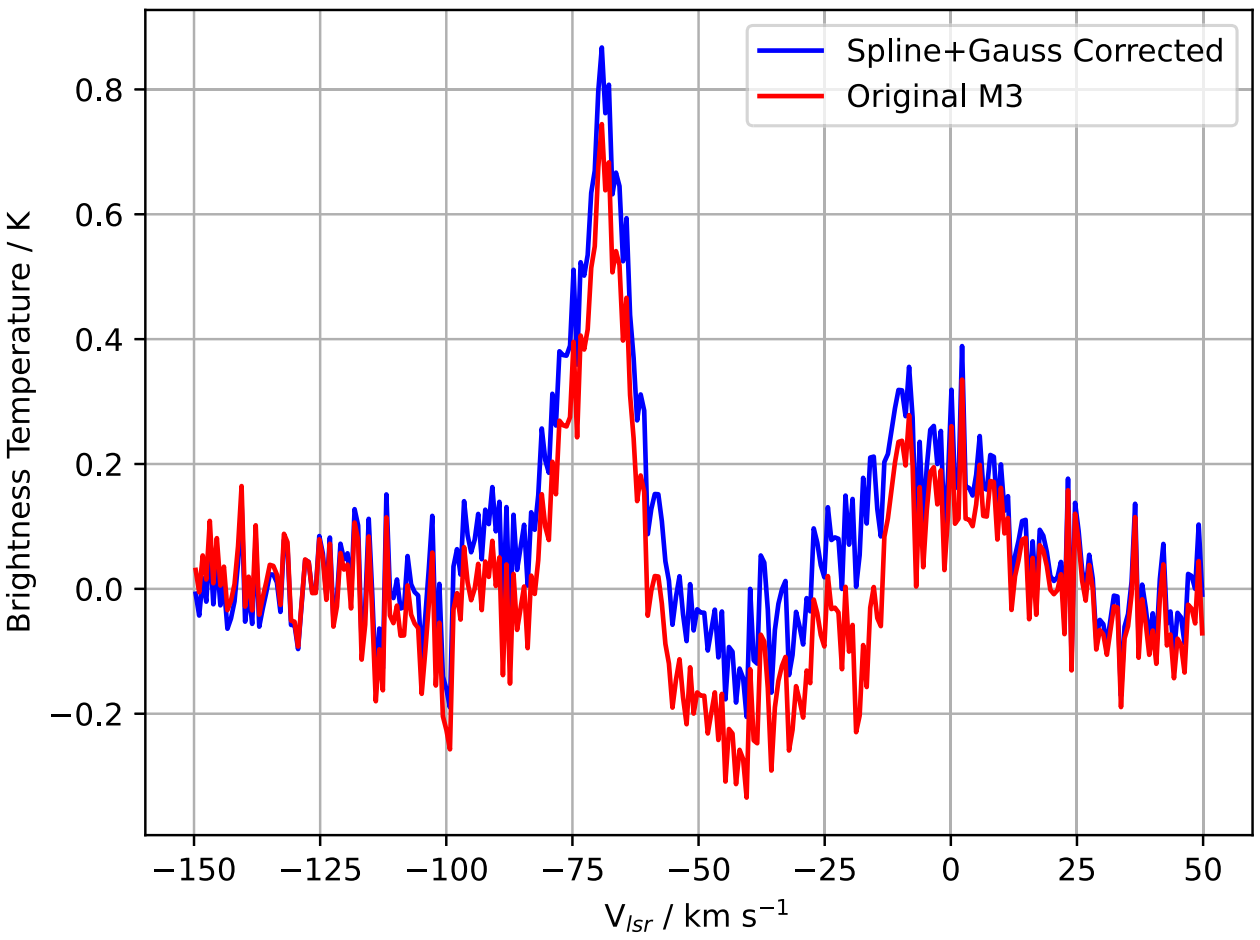

**Figure 2.** SOFIA I[C II] spatially averaged spectra across the $-150$ km s$^{-1}$ to 50 km s$^{-1}$ velocity range for the M3 region before (*red*) and after (*blue*) spline and Gaussian correction (see the text).

[1] The SOFIA project was cancelled in 2022.

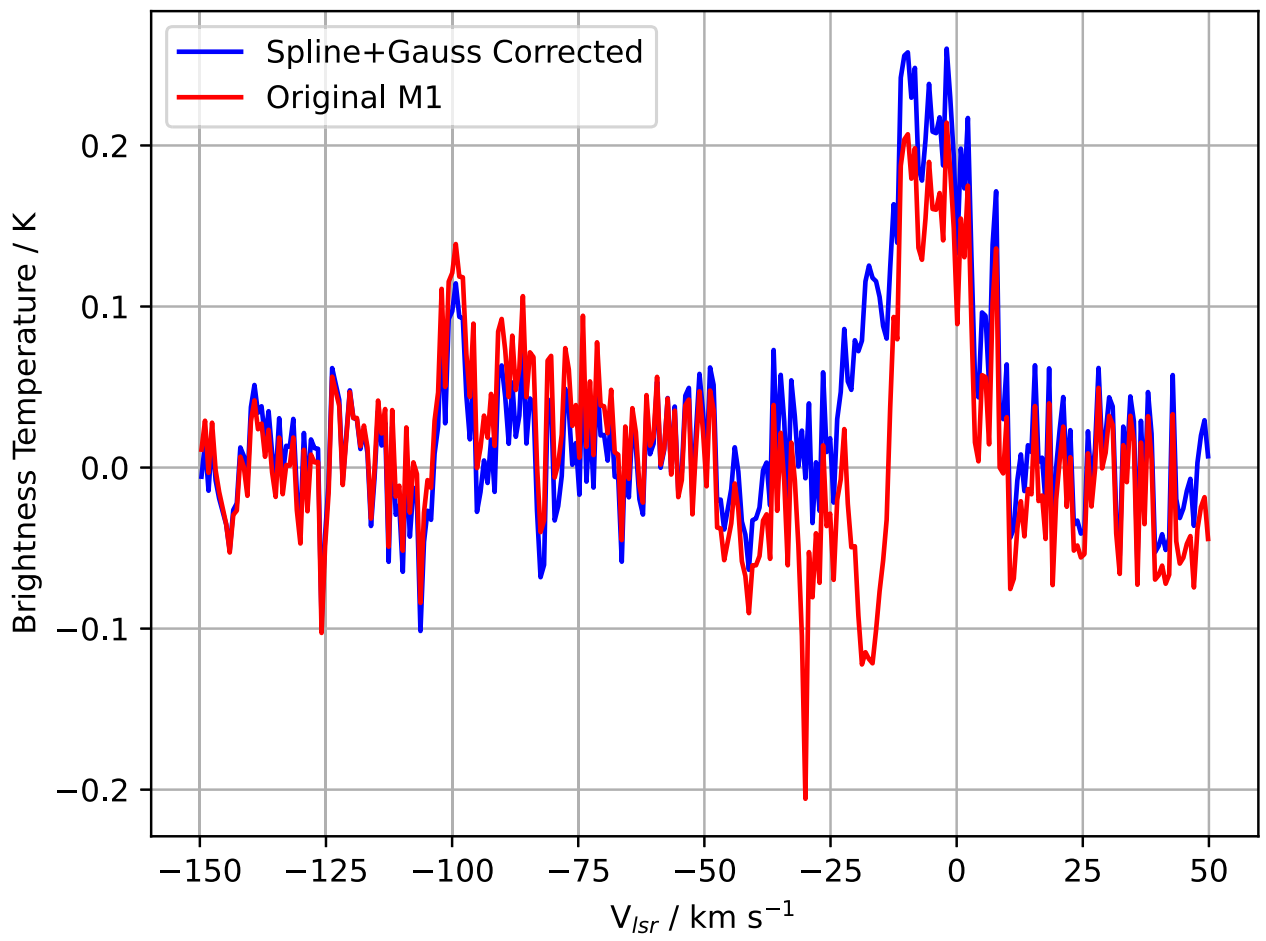

**Figure 3.** SOFIA I[C II] spatially averaged spectra across the −150 km s$^{-1}$ to 50 km s$^{-1}$ velocity range for the M1 region before (*red*) and after (*blue*) the spline and Gaussian correction (see the text).

### 2.2 Mopra [$^{12}$CO(J = 1–0)]

Mopra is a 22 m single-dish telescope near Coonabarabran, NSW, Australia (M. Burton et al. 2013; C. Braiding et al. 2018; K. Cubuk et al. 2023). The $^{12}$CO data were taken as part of the Mopra CO survey (K. Cubuk et al. 2023).

The $^{12}$CO data cover a velocity range of −150 km s$^{-1}$ to 50 km s$^{-1}$. The $^{12}$CO (J = 1–0) transition emits a photon with a frequency of 115 GHz, which is detected using Mopra's millimetre-wavelength operating band. Since $^{12}$CO is a tracer for hydrogen, we can use these data to investigate the ratio of C$^{+}$ to neutral molecular gas. The $T_{RMS}$ value is ∼ 1.2K per channel on average for Mopra $^{12}$CO data across the Mopra Southern Galactic Plane Survey, while the spectral resolution is ∼0.92 km s$^{-1}$ (C. Braiding et al. 2018; K. Cubuk et al. 2023).

### 2.3 SGPS [H I]

The H I data were taken from the Southern Galactic Plane Survey (SGPS; N. McClure-Griffiths et al. 2005). The SGPS imaged the 21 cm continuum and H I spectral line in the fourth quadrant of the Galactic plane with an angular resolution of 2.2 arcmin, using the Australian Telescope Compact Array (ATCA) and the 64 m Parkes telescope (N. McClure-Griffiths et al. 2001). The H I data have a spectral resolution of 0.8 km s$^{-1}$, a $T_{RMS}$ of ∼ 1.6 K per channel, and a velocity range of −250 km s$^{-1}$ to 150 km s$^{-1}$(N. McClure-Griffiths et al. 2005). These H I data are used to measure the ratio of C$^{+}$ to atomic gas.

## 3 SOFIA [C II] DATA PROCESSING

The SOFIA data were calibrated using the methods detailed in X. Guan et al. (2012), which covers the conversion from raw counts to antenna temperature including the correction for atmospheric transmission. The data quality is pixel-dependent with some requiring no additional post-processing while others are either discarded or baseline fitted. For more details on this process and the methods used here, see R. Higgins et al. (2021).

The baseline of the [C II] data from the M1, M2, and M3 regions was contaminated by [C II] emission in the OFF region. This was corrected before any analysis was conducted on the SOFIA data. Figs 2 and 3 show examples of the [C II] spectrum before (red) and after (blue) baseline processing. For this data set we used a univariate spline to model the baseline fit. This approach gave more control of the shape of the baseline fit (versus a polynomial fit) and made it possible to avoid contaminating the line emission where [C II] was detected (−100 km s$^{-1}$ to −75 km s$^{-1}$ and −25 km s$^{-1}$ to 0 km s$^{-1}$). For the data reduction a spline model was generated on a region with no [C II] emissions. The baseline shape is universal across regions M1, M2, and M3; therefore, the baseline fit for each sub-region (M1-1 to M1-30, M2-1 to M2-6, M3-1 to M3-5) is a scaled version of the spline baseline fit to region M1. This was done to avoid any potential bias or over-corrections in the model. The spline modelling was done for each SOFIA region and applied to the entire velocity range of −150 km s$^{-1}$ to 50 km s$^{-1}$.

The [C II] data obtained from SOFIA also had a negative dip in the spectra present at ∼−20 km s$^{-1}$. We believe this dip was the result of [C II] emission in the OFF region, which was not noticed when selecting the OFF region prior to the observations. In order to correct this dip in the spectra we applied a Gaussian model to the M3 spectra (Figs C1 and C2). Since there is little to no signal in the M3 region for that velocity range (shown by the red spectra in Fig. 2), it is the best available spectra we have to model this negative dip.

The effect of these corrections can be seen in Figs 2 and 3. There is a slight increase in the brightness temperature of the I[C II] spectra in the −25 km s$^{-1}$ to 0 km s$^{-1}$ range as a result of the spline correction.

## 4 OBSERVATIONAL ANALYSIS

In this section we will look closely at the [C II] emission from SOFIA, and how it compares to the other ISM gas tracers, $^{12}$CO from Mopra and H I from the SGPS, as well as the emission in X-rays and $\gamma$-rays. A particular focus will be on examining the profiles of [C II] against the inferred hydrogen column density from $^{12}$CO and H I.

### 4.1 [C II], $^{12}$CO, and H I across RX J1713

The ratio of I[C II] emission to total gas column density traces variations in C$^{+}$ production across the remnant. We measured I[C II], N(H I), and N(H$_2$) by averaging over latitude, longitude, and the relevant velocity range. Molecular and atomic hydrogen data were taken from Nanten and SGPS as presented in Y. Fukui et al. (2012).

Fig. 4 presents spectra from three M1 sub-regions, showing clear [C II] and $^{12}$CO peaks within the –20 to 0 km s$^{-1}$ range identified by Y. Fukui et al. (2012). Since [C II] emission extends beyond this range, we expanded the velocity interval to –25 to 0 km s$^{-1}$ for our [C II] analysis. Additional [C II] emission appears between –100 and –75 km s$^{-1}$. As we discuss later, this emission is associated with a background molecular cloud between the Norma and 3-kpc Galactic arms (Y. Moriguchi et al. 2005).

Fig. 5 shows the variation of the I[C II] to gas column density ratio across the M1 region of the remnant. The atomic and molecular gas data were taken from Y. Fukui et al. (2012; SGPS and Nanten, respectively). The self-absorption of the H I atomic emission was corrected for by Y. Fukui et al. (2012). The ratio of I[C II] to the total gas column density is shown in Fig. D1.

We use the integrated intensity of [C II] directly, avoiding conversion to N($C^+$) due to the large uncertainties introduced by assumptions about local electron temperature and density (P. Goldsmith et al. 2012).

In the middle panel of Fig. 5, [C II] anticorrelates with $H_2$, indicating that [C II] is more closely associated with atomic gas. While the molecular gas varies significantly across M1, the atomic gas remains relatively constant, with the total column density (Fig. D1) mirroring the molecular gas profile.

The X-ray and TeV $\gamma$-ray profiles peak towards the edge of the remnant (close to the SNR shock) and are at their lowest at the centre of the remnant. The H.E.S.S. TeV $\gamma$-ray peak near $l \sim 347.5^\circ$ aligns with a rise in the I[C II]/N(H I) ratio. There is no corresponding X-ray peak at $l \sim 347.5^\circ$. Both the X-ray and TeV $\gamma$-ray peak on the right side of the remnant, corresponding to the peak of the N($H_2$) in the middle panel. The *Fermi*-LAT GeV profile stays fairly flat across the SNR. This is likely due to its resolution ($\sim 0.1^\circ$), which leads to less detail in its profile as compared to the H.E.S.S. TeV and X-ray keV.

$H_2$ column densities are roughly twice those of H I near $l \sim 347^\circ$, corresponding to the molecular clump known as Core C (Y. Fukui et al. 2012; N. I. Maxted et al. 2013). Elsewhere, atomic and molecular column densities are comparable.

For completeness, we include a plot of the spatial profiles of I[C II], N(H I), N($H_2$), X-ray and $\gamma$-ray emissions across M1 (Fig. B1). The H I column density is relatively uniform ($\sim$ 20 per cent variation), while I[C II] peaks on the far left side, outside the SNR shell (shown by the black dashed lines). The [C II] and $^{12}$CO intensities rise sharply on the SNR's eastern edge.

Figs E1–E3 show channel maps of I[C II] across RX J1713. In Fig. E1, [C II] emission from –14 to –8 km s$^{-1}$ spatially aligns with $^{12}$CO, suggesting these clouds contain ionized components.

In summary, across M1, we find that the [C II] follows the atomic gas closely on the left side of the remnant, where we also see a peak in the TeV $\gamma$-ray profile around $l \sim 347.5^\circ$. On the right side of the remnant, we see a peak in both the keV X-ray and TeV $\gamma$-ray at $l \sim 347.0^\circ$. However, this does not follow the I[C II]/N($H_2$) ratio, which suggests the [C II] does not follow the molecular gas.

### 4.2 Region M2

The M2 region, located southeast of the SNR, contains dense, self-absorbed atomic gas, which appears to overlap TeV $\gamma$-ray emission (Y. Fukui et al. 2012), and shows minimal X-ray emission. Fig. 6 presents the [C II] and $^{12}$CO spectra averaged over the region. There is a clear $^{12}$CO peak at $-25$ km s$^{-1}$ indicating molecular gas behind the SNR, possibly at the edge of the SNR shell. Compared to M1, X-ray emission in M2 is significantly weaker. However, nearby H.E.S.S. $\gamma$-ray emission suggests the presence of cosmic-ray protons.

Fig. 7 shows that the I[C II]/N($H_2$) ratio reaches up to $\sim 25 \times 10^{-21}$ K km s$^{-1}$ / cm$^{-2}$ due to the low molecular gas content, while the I[C II]/N(H I) ratio remains much lower ($\sim 0.4 \times 10^{-21}$ K km s$^{-1}$ / cm$^{-2}$).

The I[C II] emission is concentrated between –25 and –10 km s$^{-1}$ and spatially overlaps with H I, as shown in Fig. E2,

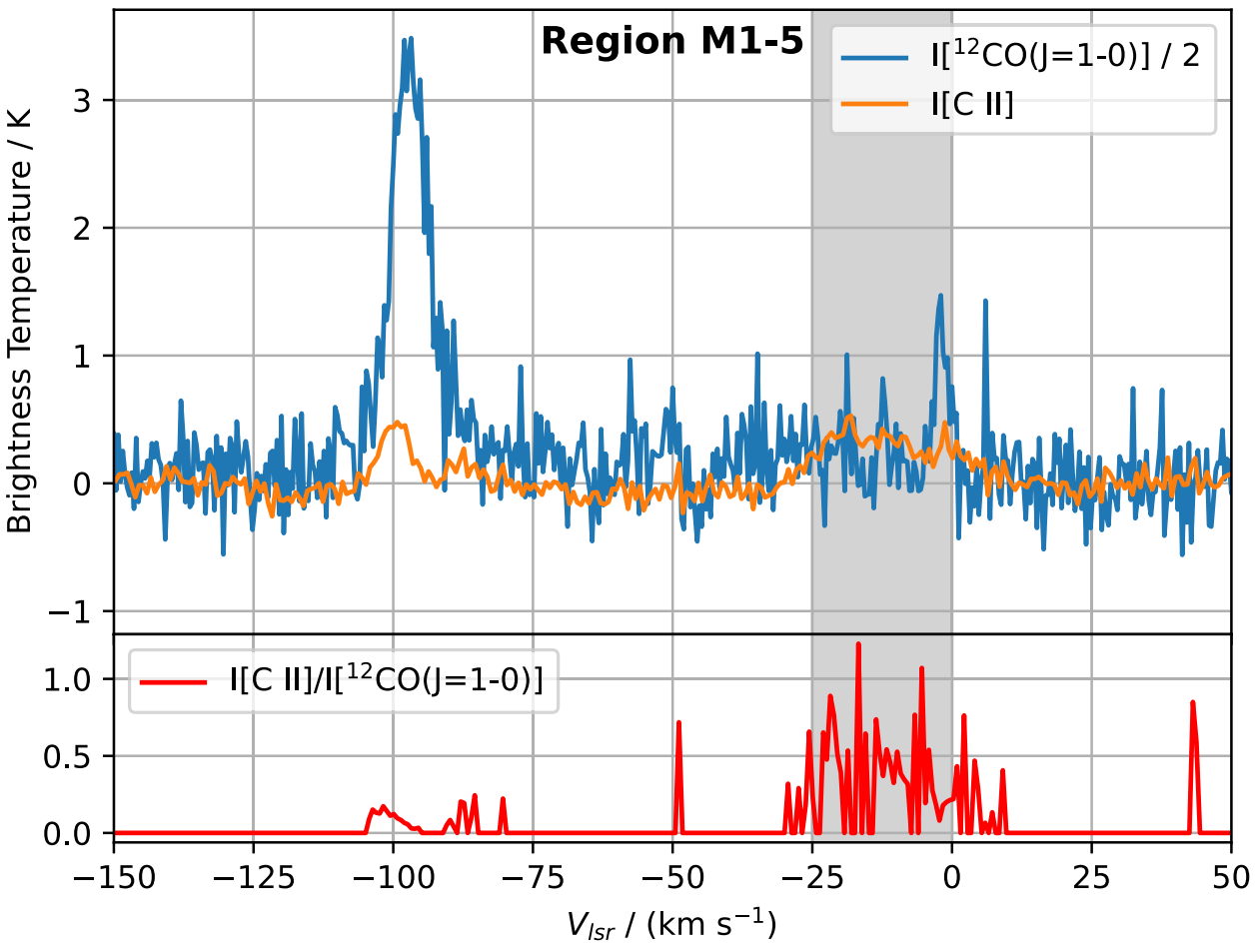

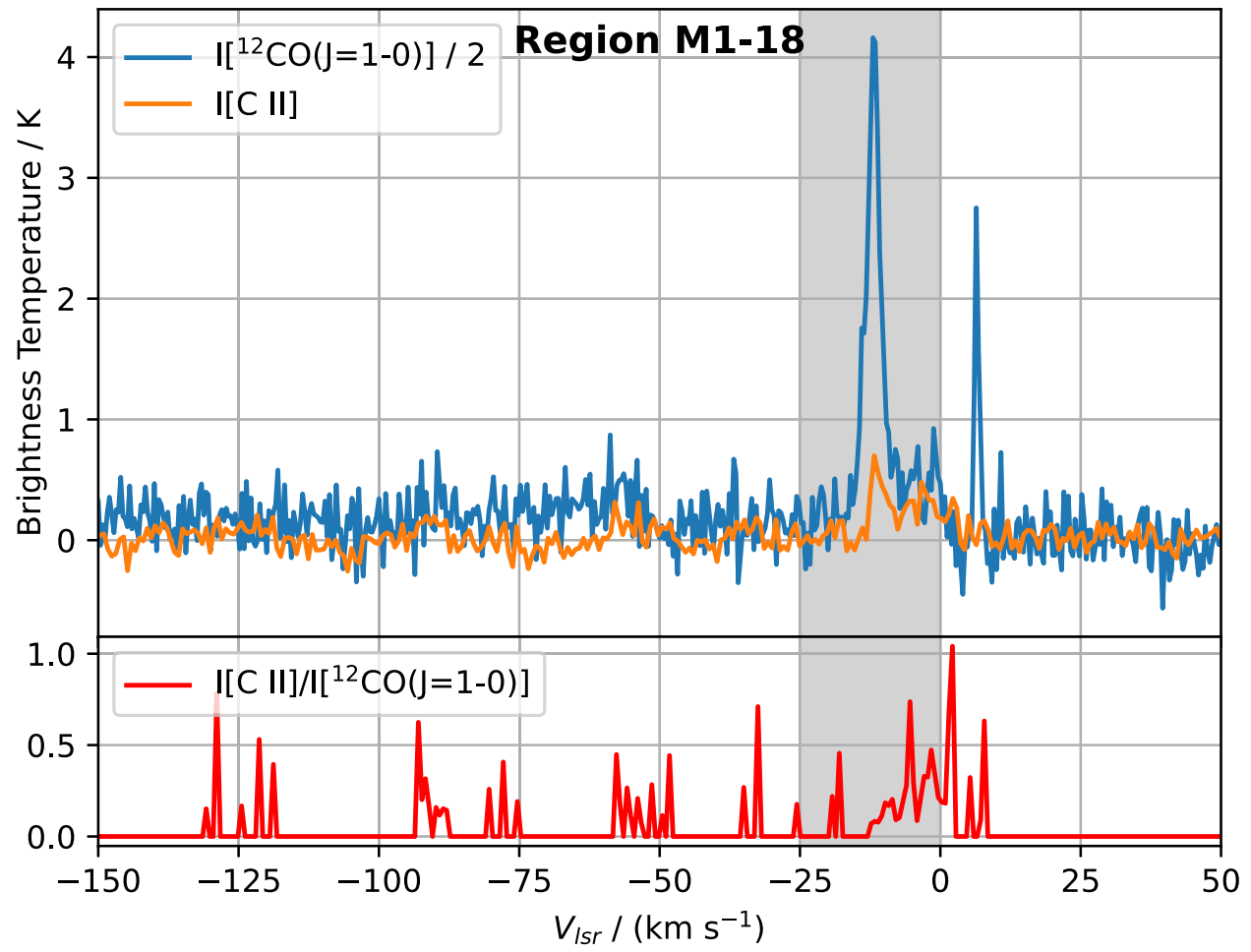

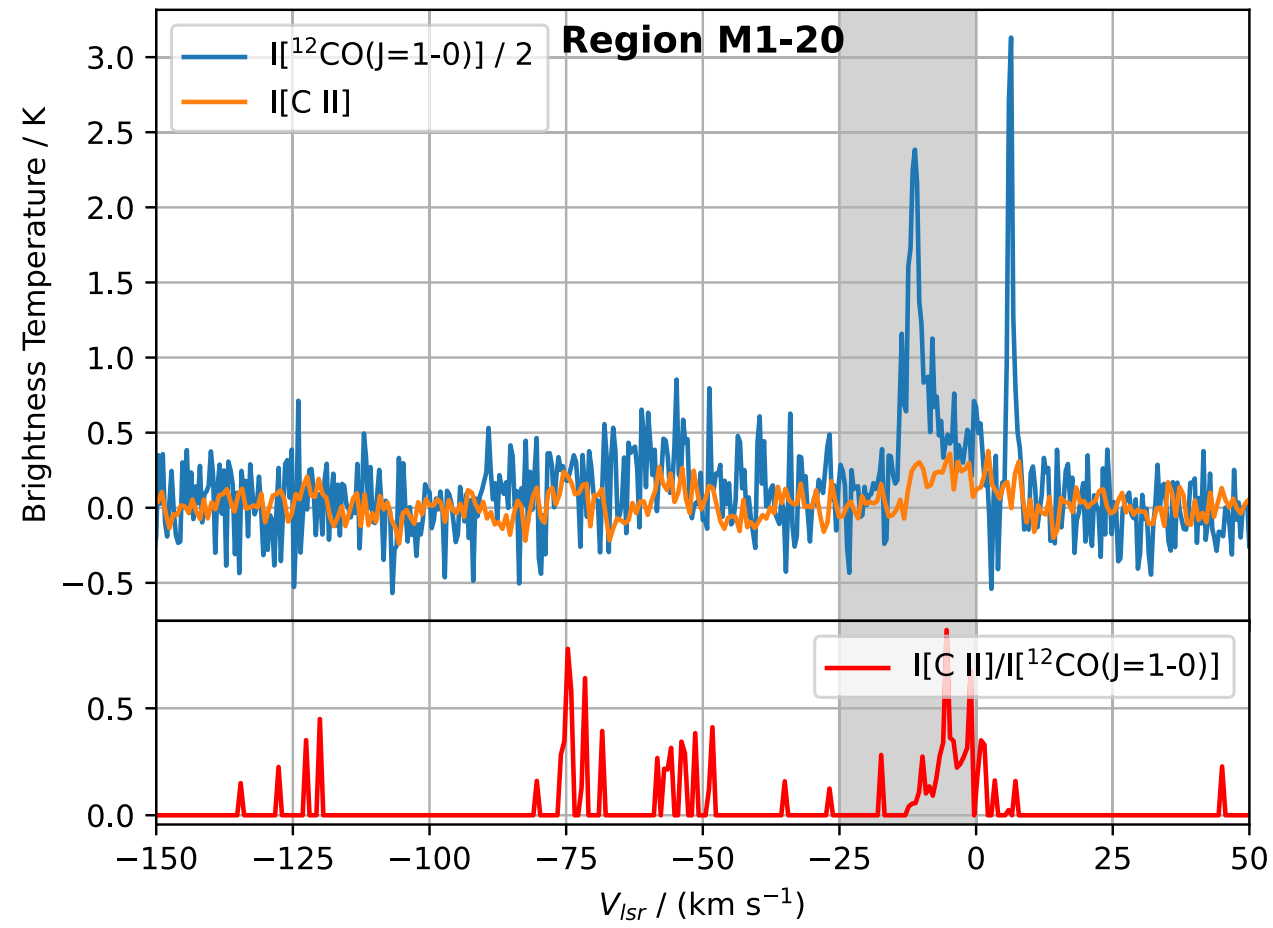

**Figure 4.** Mopra I[$^{12}$CO(J = 1–0)], SOFIA I[C II] spectra and their ratio from selected M1 regions. The [C II] and $^{12}$CO spectra are averaged over the M1 region, and I[$^{12}$CO] is divided by 2 for better comparison with I[C II]. A running mean has been applied to the I[C II] and I[$^{12}$CO] spectra for visual clarity. Full resolution versions are in Figs F1 and F2. The shaded band identifies the velocity range in which SNR RX J1713 is located.

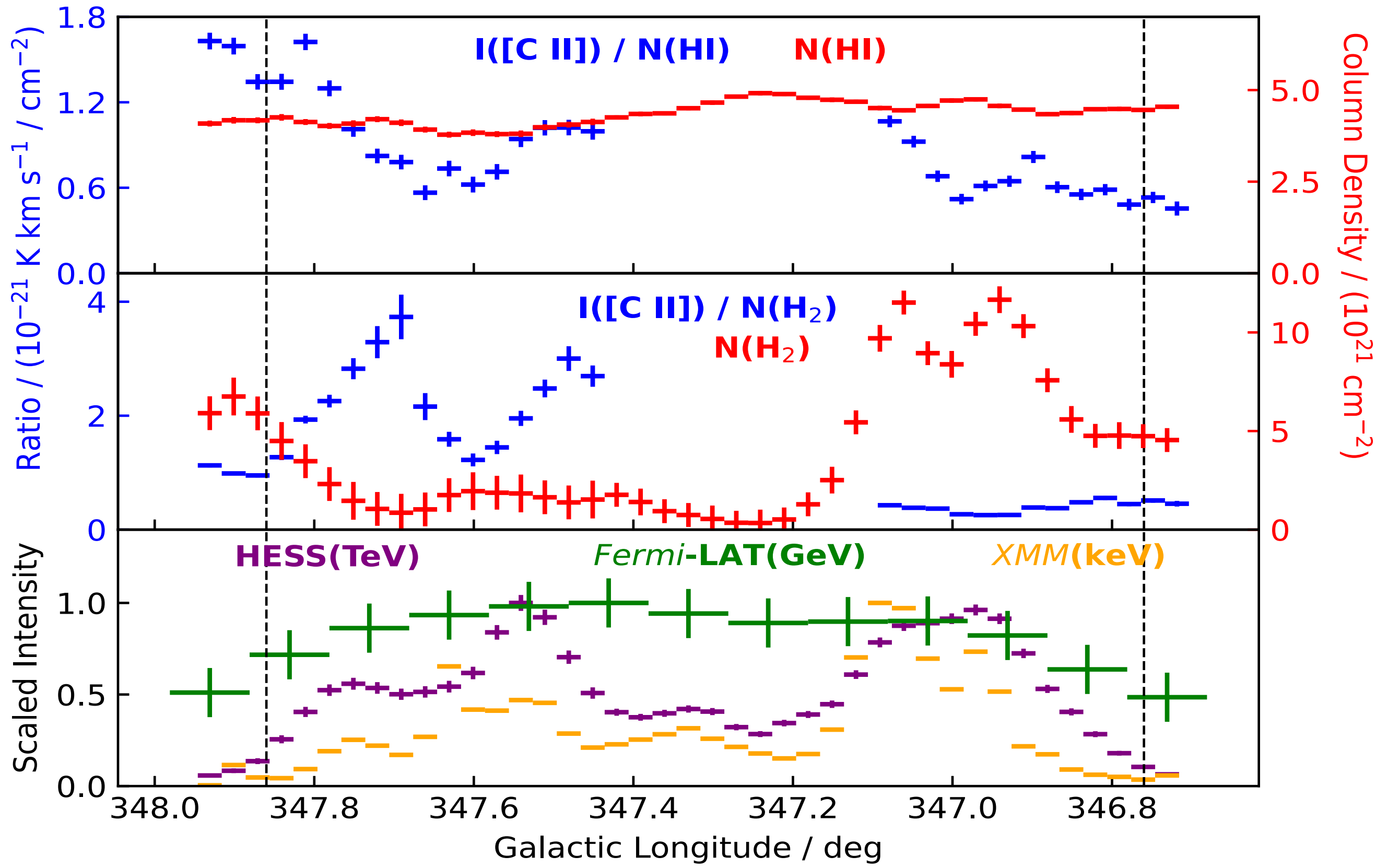

**Figure 5.** Longitude profiles of the ratio of integrated I[C II] to the atomic (*top*) and molecular (*middle*) gas column densities across region M1 of the SNR RX J1713 (*blue*), and the variation in the column densities of atomic and molecular gas (*red*). The scales and labels in red refer to the column densities of N(H I) and N($H_2$), while the blue labels refer to the ratio of I[C II] to the column densities. The black dashed lines show the outer edge of SNR RX J1713, based on the *XMM–Newton* X-ray image. The spectra were integrated across the velocity range of $-20\,\mathrm{km\,s^{-1}}$ to $0\,\mathrm{km\,s^{-1}}$. *Bottom panel*: The profile of H.E.S.S. $\gamma$-rays (> 2 TeV), *XMM–Newton* X-rays, and *Fermi*-LAT $\gamma$-rays (500 MeV–300 GeV), scaled to their maximum values, across the remnant over the same regions as the SOFIA [C II] M1 mapping.

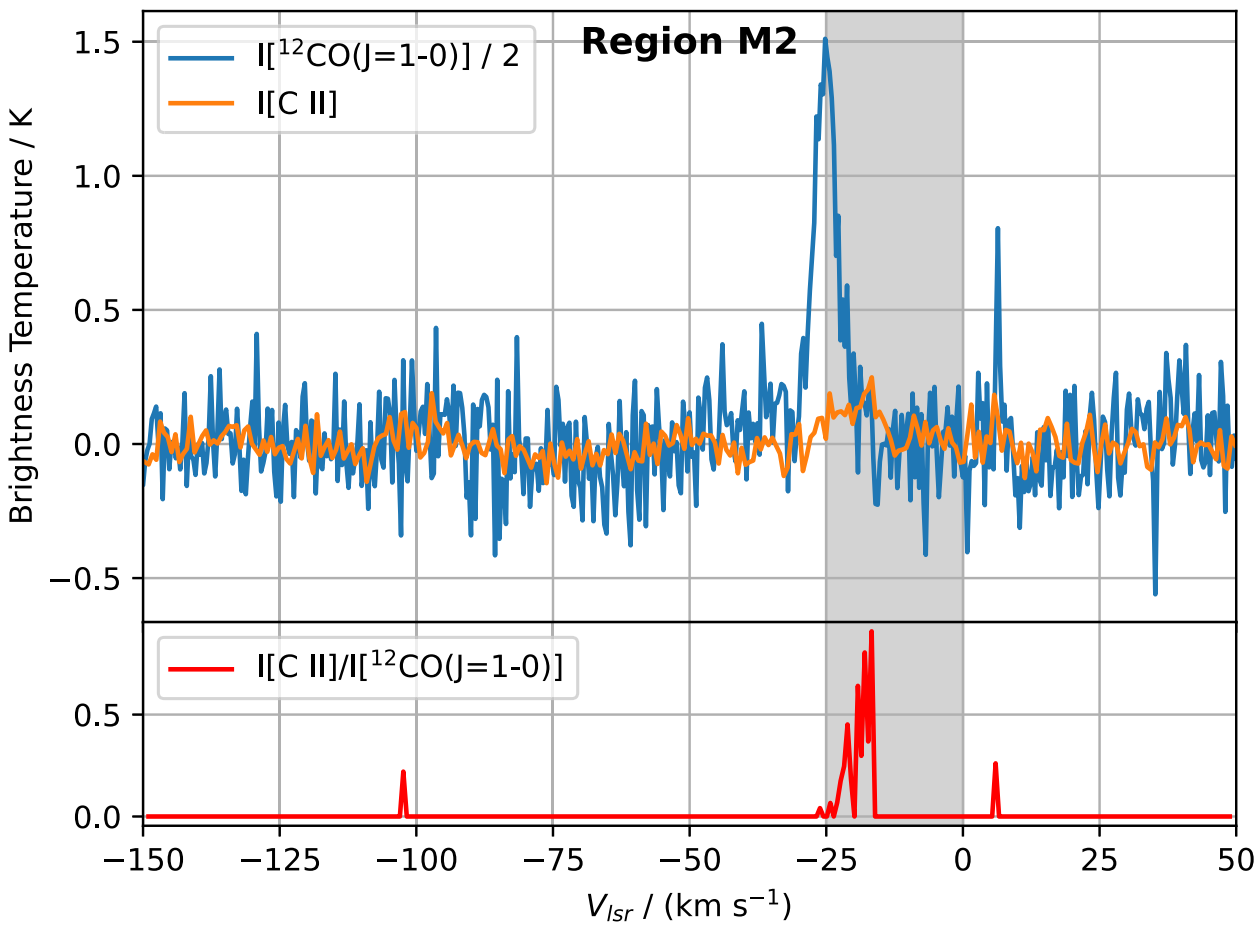

**Figure 6.** Mopra I[$^{12}$CO(J = 1–0)], SOFIA I[C II] and their ratio spectra from the M2 region. The $C^+$ and $^{12}$CO spectra are averaged over M2, and $^{12}$CO is divided by 2 for better comparison with $C^+$. The shaded band identifies the velocity range in which SNR RX J1713 is located.

indicating that $C^+$ is primarily associated with the atomic gas in this region.

### 4.3 Region M3

The M3 region lies along the northern shock front of the SNR and coincides with peaks in both X-ray and $\gamma$-ray emission. N. I. Maxted et al. (2013) and H. Sano et al. (2020) found dense molecular gas clumps and filaments in the region, highlighting the complex spatial structures of the molecular gas in this region. Fig. 8 shows strong I[$^{12}$CO] and I[C II] emission within the relevant velocity range, with the I[C II] peak $\sim$35 km s$^{-1}$ wide. An additional [C II] peak is observed at $\sim$ - 75 km s$^{-1}$, which is associated with the molecular cloud in the background mentioned in the previous section.

In M3, spatial variation in TeV $\gamma$-ray emission across latitude (Fig. 1) motivated the use of a latitude profile in Fig. 9. In addition to this, the keV X-ray emission in Fig. 1 shows the SNR shock to be running along longitude, which suggests that a latitude profile may provide more information on the dynamics in the region. Located at the northern shock front (Fig. 1), M3 offers some insight into $C^+$ distribution within the environment of the shock.

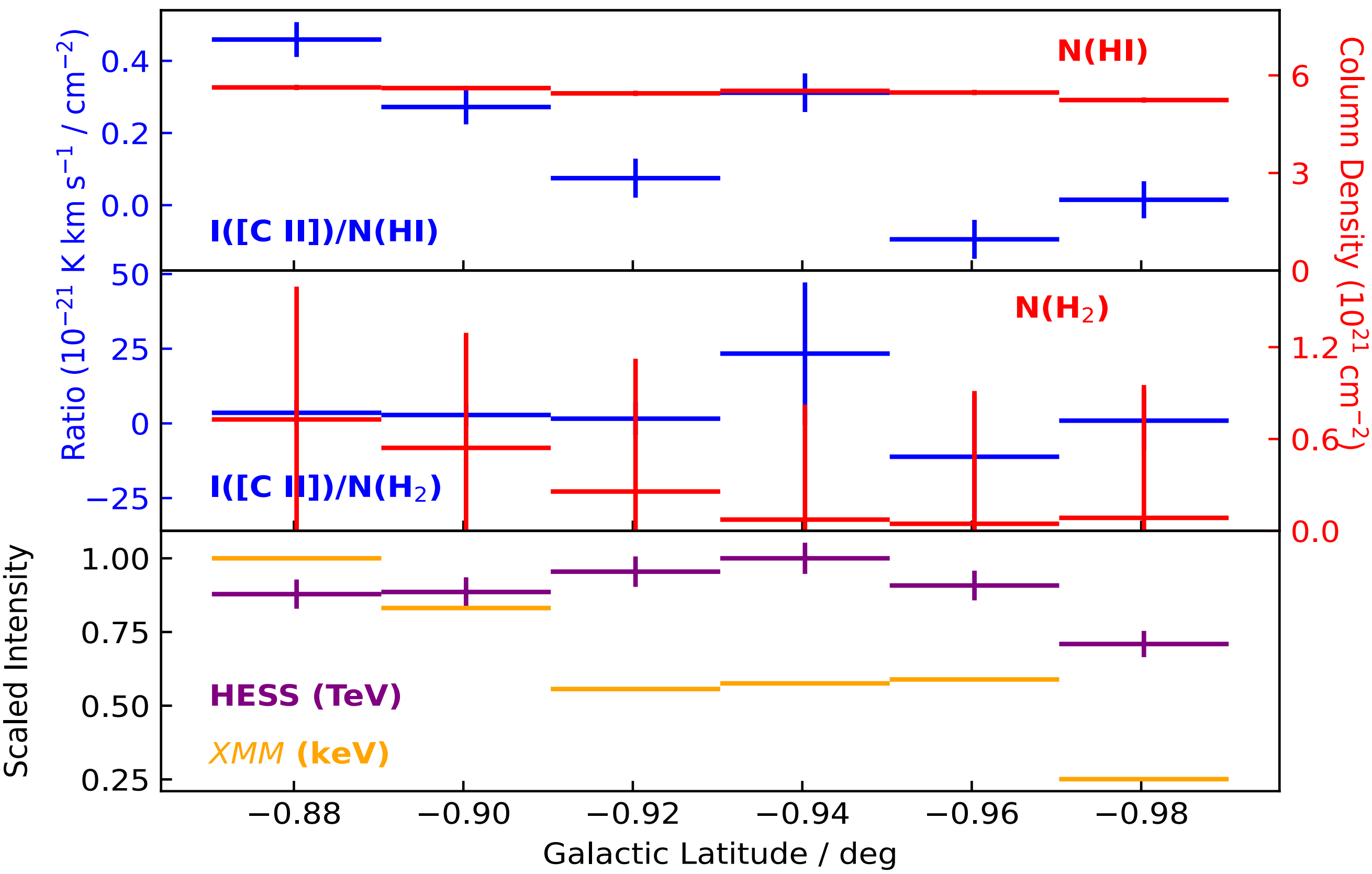

**Figure 7.** *Upper panel*: Latitude profiles of the ratio of the I[C II] to atomic and molecular gas column density across the region M2 of the SNR and the velocity range of $-20$ km s$^{-1}$ to 0 km s$^{-1}$. *Lower panel*: The profile of H.E.S.S. $\gamma$-rays (> 2 TeV) and *XMM–Newton* X-rays, scaled to their maximum values, across the remnant over the same region as the SOFIA M2 mapping.

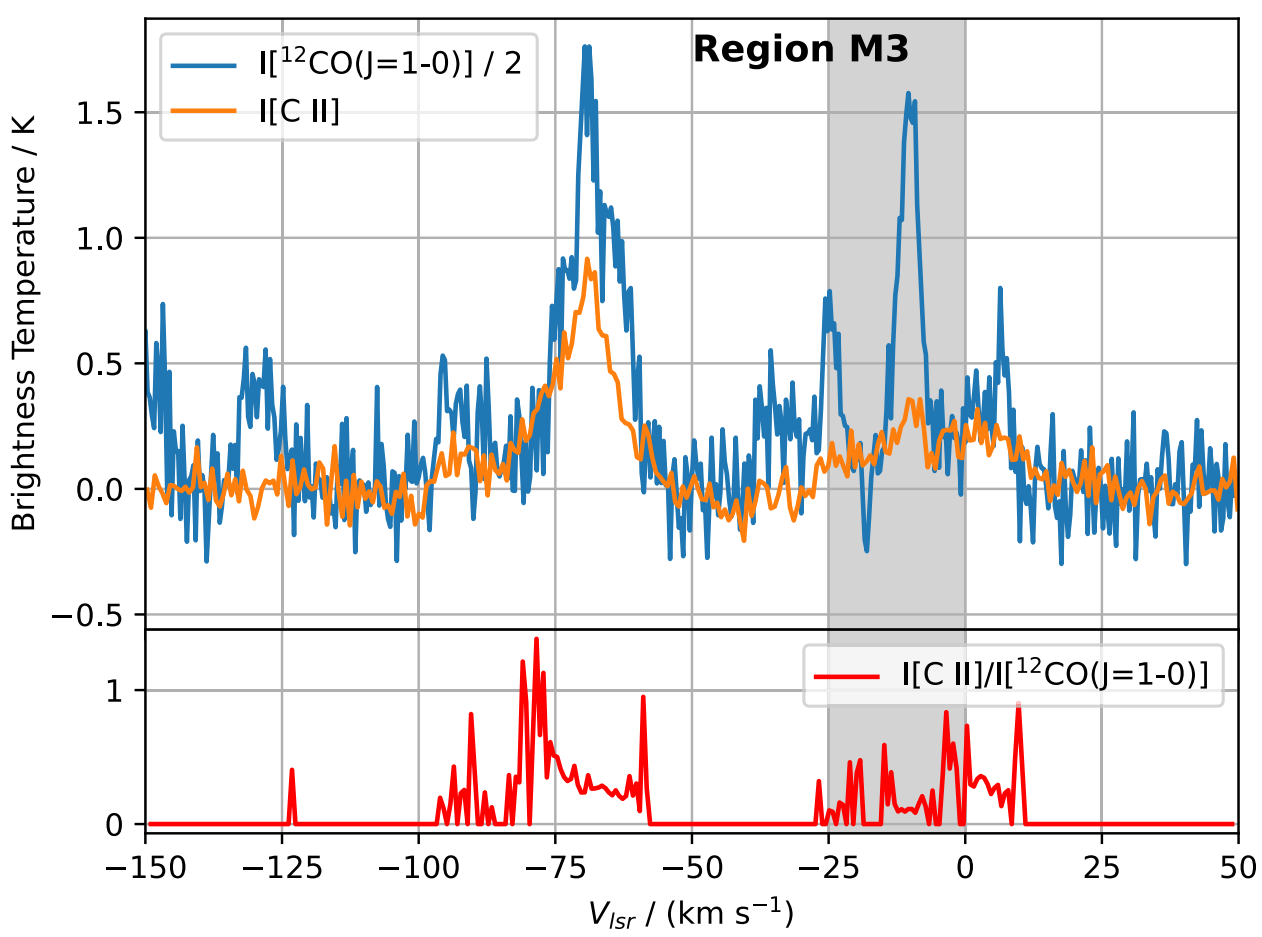

**Figure 8.** Mopra I[$^{12}$CO(J = 1–0)], SOFIA I[C II] spectra and their ratio from the M3 region. The $C^+$ and $^{12}$CO spectra are averaged over M3, and $^{12}$CO is divided by 2 for better comparison with $C^+$. The shaded band identifies the velocity range in which SNR RX J1713 is located.

The I[C II]/N(H I) and I[C II]/N($H_2$) ratios reach $\sim$1.2 and $\sim$1.1 $\times 10^{-21}$ K km s$^{-1}$ / cm$^{-2}$, respectively. This suggests that $C^+$ is distributed evenly across both atomic and molecular gas.

Fig. E3 shows that I[C II] emission follows $^{12}$CO contours between –15 and 0 km s$^{-1}$. This is associated with the background molecular gas cloud mentioned in the previous section.

## 5 [C II] IN THE GOT C+ SURVEY

To put the [C II] emission towards SNR RX J1713 into a wider context, we examined the pointings from the *Herschel* GOT C+ survey (T. Velusamy et al. 2010; J. Pineda et al. 2013; W. Langer et al. 2014) plus their corresponding Mopra $^{12}$CO and SGPS H I observations. The GOT C+ survey comprised 394 pointings (T. Velusamy et al. 2010; J. Pineda et al. 2013; W. Langer et al. 2014), positioned across the Galactic plane in a grid arrangement. These pointings therefore somewhat randomly covered star formation regions, supernova remnants and other objects of interest.

From W. Langer et al. (2014), we have an average $T_{\rm RMS}$ of 0.08 K km s$^{-1}$ and a minimum $T_{\rm RMS}$ of 0.06 K km s$^{-1}$ for GOT C+. The $T_{\rm RMS}$ for the SOFIA data in the $-25$ km s$^{-1}$ to 0 km s$^{-1}$ range is 0.008 K km s$^{-1}$. These values would suggest SOFIA is significantly more sensitive than *Herschel* and is probing a lower level of I[C II] than GOT C+. Although the performance of *Herschel* HIFI and SOFIA upGREAT are similar (R. Higgins et al. 2021), the SOFIA observations for RX J1713 were averaged over a larger area and as a result have a longer integration time (1–2 s for GOT C+ and 500 s for SOFIA), which explains the lower $T_{\rm RMS}$ for the SOFIA data (R. Higgins et al. 2021). The variations in observing conditions between SOFIA and *Herschel* make it difficult to draw detailed conclusions by comparing the observations. For completeness, the histograms comparing the SOFIA and GOT C+ data sets are shown in Section G.

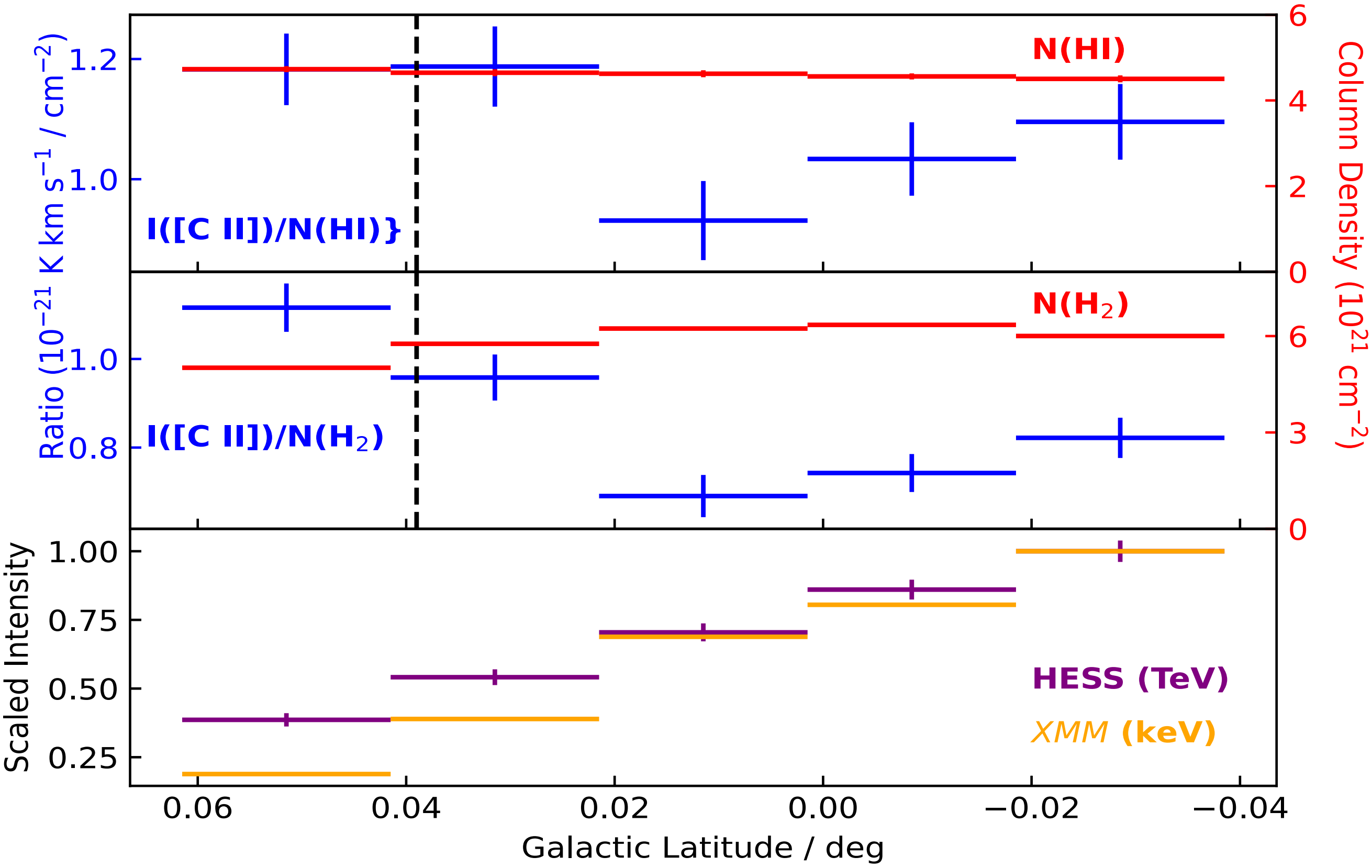

**Figure 9.** *Upper panel*: Latitude profiles of the I[C II]/N(H I) and I[C II]/N(H$_2$) ratio across region M3 of SNR RX J1713 and a velocity range of −20 km s$^{-1}$ to 0 km s$^{-1}$. *Lower panel*: The profile of H.E.S.S. $\gamma$-rays (> 2 TeV) and *XMM–Newton* X-rays, scaled to their maximum values, across the SNR over the same region as the I[C II] M3 mapping.

Since the [C II] emission is expected to vary according to ionization conditions, we separated the GOT C+ results based on their proximity to SNRs and H II regions to determine how the I[C II]/I[$^{12}$CO] ratio varies for each one. A GOT C+ pointing was considered to be associated with an SNR or an H II region if it was located within 0.1deg and 10 km s$^{-1}$ of these objects. These were chosen based on the typical size of H II regions (G. Garay & S. Lizano 1999) and SNRs (S. Ranasinghe & D. Leahy 2023). We used Green's SNR catalogue (D. Green 2022) and the *WISE* H II catalogue (L. Anderson et al. 2014) to find objects that would coincide with the GOT C+ pointings. If a pointing was not associated with either an SNR or H II region, it was classed as 'Other'. The ratio distributions of these subsets are shown in Fig. 10.

The dashed line in Fig. 10 indicates the median of the I[C II]/I[$^{12}$CO] ratio for the SNR and H II subsets. We adjusted the criteria for an SNR or H II region to be associated with a GOT C+ pointing over a wider range (0.05, 0.2deg and 5, 20 km s$^{-1}$) to test the stability of the subset. There is no statistically significant difference in the medians of the SNR and H II distributions in Fig. 10. Overall, we do not see a significant difference in the I[C II]/I[$^{12}$CO] ratio between the subsets of pointings close to SNRs and H II regions, and also in the rest of the pointings (labelled 'Other') in the GOT C+ survey. The green and orange dashed lines are the medians of the SOFIA histograms from Fig. G3.

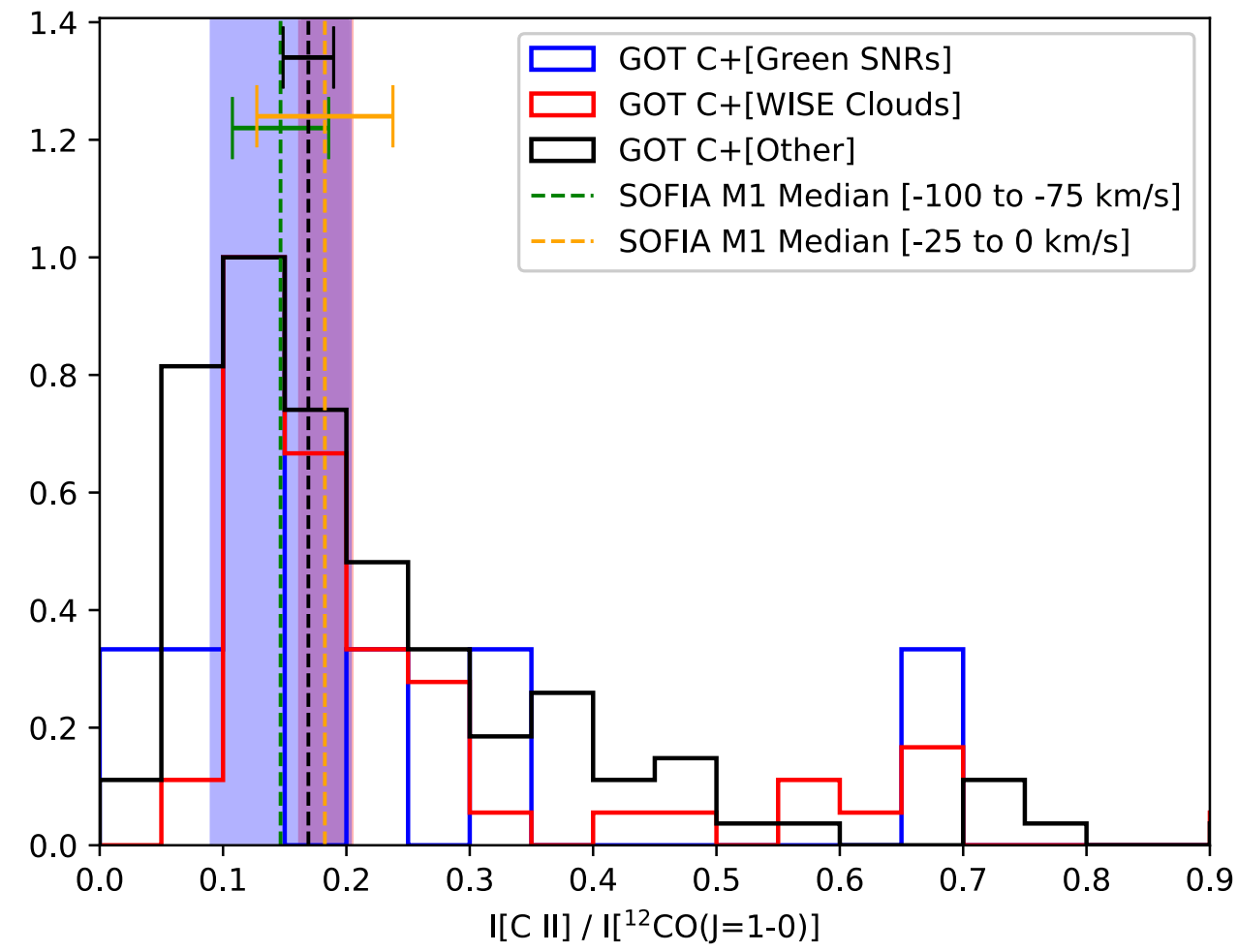

**Figure 10.** The subset of I[C II]/I[$^{12}$CO(J = 1–0)] ratios from GOT C+ which spatially coincide with SNRs (from Green's SNR catalogue), H II regions (from the *WISE* H II catalogue) and 'Other' pointings which do not coincide with an object in either catalogue. The ratio distributions are scaled to their maximum value. The vertical dashed lines indicate the median of the distribution with the corresponding colour and the error bars indicate the standard error in the median.

## 6 MODELLING OF UV EMISSION AND CR IONIZATION

The results from the previous section show there is no significant difference in the I[C II]/I[$^{12}$CO] ratio at SNRs and other regions (such as H II regions, gas clouds) in the ISM. These results suggest that the ionization of carbon is not necessarily driven by low-energy CRs from the SNR, which is in line with the notion that carbon is primarily ionized by FUV photons (A. G. G. M. Tielens & D. Hollenbach 1985). However, to investigate the contribution of UV photons to the ionization of carbon towards SNR RX J1713, we modelled the UV emission from potential sources close to the SNR. Dissociative shocks in older SNRs (> 10 000 yr), where blast wave velocities have decelerated to 20–40 km s$^{-1}$, can produce UV photons in the Lyman and Werner bands that subsequently ionize carbon and other species in surrounding neutral gas (D. J. Hollenbach, D. F. Chernoff & C. F. McKee 1989; P. Lesaffre et al. 2013; D. R. Flower & G. Pineau des Forêts 2003). Observationally, direct detection of UV emission from SNR shocks remains rare. R. A. Fesen et al. (2021) used *GALEX* FUV imaging to confirm shock-driven UV filaments in three old large, high-latitude remnants, the Antlia SNR, G249+24 and G354–33. For a young SNR like RX J1713 (∼ 1600 yr), the shock velocities remain too high for the slow dissociative-shock mechanism to contribute significantly, so we do not include this UV source in our modelling.

The photoionization code, CLOUDY (G. J. Ferland et al. 2017), was used to model the UV emission of nearby stars and H II regions, and the photoionization production of [C II] emission they might produce at SNR RX J1713. CLOUDY was also used to predict the [C II] emission assuming only CRs were ionizing the gas at RX J1713.

G. J. Ferland et al. (1998) and G. J. Ferland et al. (2017) describe the previous two major reviews of the code. G. J. Ferland (2003) and D. Osterbrock & G. Ferland (2006) explain the physics simulated by the code. CLOUDY simulates micro-physical processes that occur in the interstellar plasma and neutral medium. These include ionization (such as photoionization, collisional ionization and the Auger effect) and recombination processes (such as radiative and charge transfer). Since no analytical solutions are possible for these processes, numerical simulations are required. CLOUDY predicts the observed spectra based on radiation propagating through varying, depth-dependent physical conditions.

Another often-used code used to model ionization in gas clouds is Meudon PDR (F. Le Petit et al. 2006). This code models photon-dominated regions (PDRs), with detailed chemistry (∼ 2000 reactions with ∼ 200 species) and radiative transfer in diffuse and dense clouds. However, for our application, CLOUDY offers flexibility (variable star position, cloud distribution and UV sources) in specifying stellar radiation sources and geometric configurations. CLOUDY allows us to model individual O-type and B-type stars as distinct ionization sources around each SNR. This flexibility is essential for predicting the contributions from UV sources and CRs to [C II] emission. M. Röllig et al. (2007) compared CLOUDY, Meudon and other PDR codes. Their results suggest these PDR codes have similar predictions (∼ 15 per cent variation) for number densities, photoionization rates and emissivities when running their benchmark models. We note that our models assume a constant hydrogen density within each run rather than a non-uniform density structure. Given the large spatial scales involved (∼ 100 pc between the UV sources and the SNR), the true density profile of the intervening ISM is poorly constrained, so adopting a detailed non-uniform structure would introduce a large number of assumptions. Instead, we ran models spanning a range of constant densities (1–100 cm$^{-3}$; Table 2), which bracket the plausible conditions and achieve the objective of bounding the predicted [C II] emissivity. A more detailed treatment incorporating non-uniform density structure could be explored in future work, particularly for SNRs where the surrounding ISM is better constrained.

**Table 2.** The range of parameters for the CLOUDY models. The table shows the different values used for hydrogen density, distance of O9.5 stars, distance to H II regions and CR ionization rate.

| Hydrogen density (cm$^{-3}$) | O9.5 stars (pc) | H II regions (pc) | CR ionization rate (s$^{-1}$) |
|---|---|---|---|
| 1 | 85 | 34 | $10^{-14}$ |
| 10 | 100 | 40 | $10^{-15}$ |
| 100 | 115 | 46 | $10^{-16}$ |

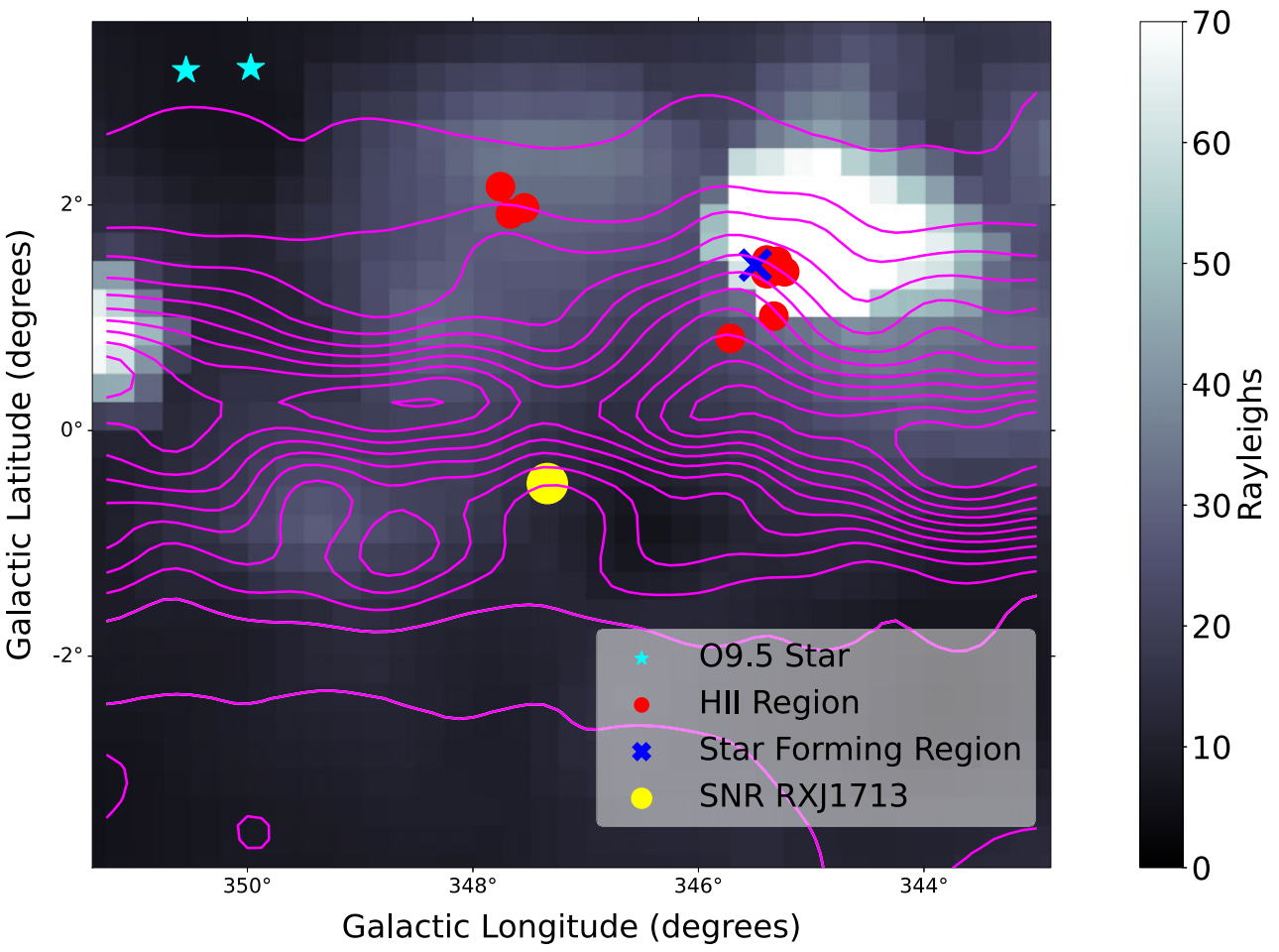

**Figure 11.** H $\alpha$ image (Rayleigh units) taken from the WHAM survey (L. M. Haffner et al. 2010), integrated from −25 km s$^{-1}$ to 0 km s$^{-1}$, in the region surrounding SNR RX J1713. The purple contours are $^{12}$CO emission, integrated from −25 km s$^{-1}$ to 0 km s$^{-1}$ from Mopra (C. Braiding et al. 2018; K. Cubuk et al. 2023). The yellow dot shows the location of SNR RX J1713 based on the X-ray and gamma-ray emission (G. Cassam-Chennai et al. 2004; H.E.S.S. Collaboration 2018b). The other markers show the positions of O9.5 stars (Gaia Collaboration 2020; cyan stars), a star-forming region (M. F. Skrutskie et al. 2006; blue cross) and H II regions from the *WISE* catalogue (L. Anderson et al. 2014; red dots).

H $\alpha$ emission is an indicator of the ionization of hydrogen, a tracer for star formation and the presence of young, massive stars (with significant UV emission). Fig. 11 shows the distribution of H $\alpha$ covering the RX J1713 region (L. M. Haffner et al. 2010). The markers in Fig. 11 show O9/9.5 stars, star-forming regions and H II regions in the same velocity range as the SNR. As we can see, the likely prominent sources of UV photons are located to the north of SNR RX J1713. The UV emission from these sources must propagate through the ISM before reaching the SNR. Our CLOUDY model assumes UV photons are produced by a source (O9.5 star, H II region) which is surrounded by a ring of ISM gas. The distance from the UV source to the ISM ring and the thickness of the ISM ring itself are determined by the H $\alpha$ distribution in Fig. 11. Fig. 12 shows a simplified schematic of the model used in CLOUDY. We define the depth into the cloud, $R_c$, as the distance measured from the inner face of the ISM ring (illuminated by the UV source) to its outer edge, which is located at SNR RX J1713; CRs are assumed to enter from this outer edge.

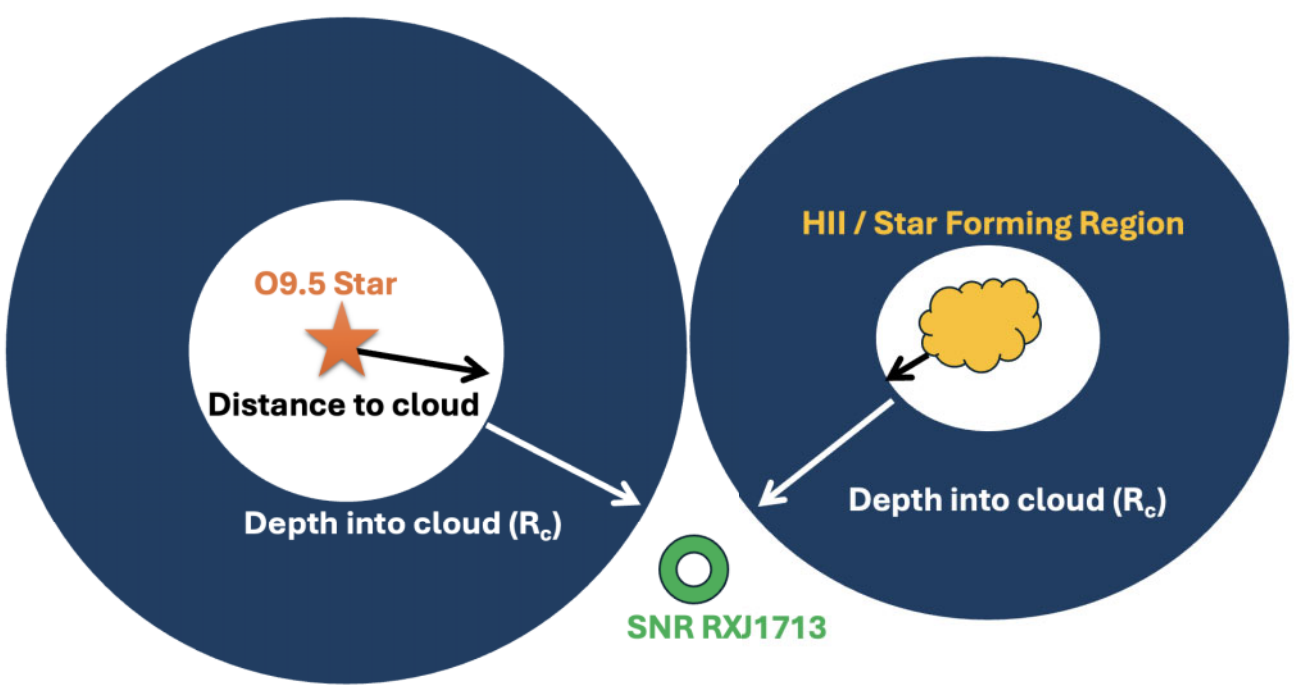

**Figure 12.** A schematic drawing of the scenario surrounding RX J1713 for our CLOUDY modelling. Each star and H II/star-forming region is assumed to be at the centre of a ring of ISM gas, the outer edge of which is located at SNR RX J1713. The UV emission from the star or H II region impacts the inner face of the ISM ring and cascades across the depth of this ring. The depth $R_c$ within the ring is defined as the distance starting from the inner face of the ring to its outer edge.

As Fig. 11 shows, there are three groups of objects at distances consistent with $-25\,\mathrm{km\,s^{-1}}$ and $0\,\mathrm{km\,s^{-1}}$. The three groups are the two O9.5 stars in the top left, the three H II regions directly above the SNR and the nine H II regions and one star-forming region towards the top right in Fig. 11. The two O9.5 stars (top left of Fig. 11) are located slightly above the galactic plane ($\sim 100$ pc away from the SNR). There is also a noticeable void in the H $\alpha$ emission around these stars. Based on the size of these voids, their UV emission is assumed to travel $\sim 30$ pc before it interacts with ISM gas, based on the position in Fig. 11. The two other groups are located within ISM clouds.

The UV emission from the objects shown in Fig. 11 was modelled using CLOUDY to output the [C II] emission they produce at SNR RX J1713. This [C II] emissivity was then compared to the [C II] emissivity from CRs, modelled using CLOUDY. The distance between the two groups of H II regions and the SNR was set at 40 pc. Since the two groups are embedded within gas clouds, the UV emission was assumed to be attenuated by this ISM gas after travelling $< 1$ pc.

The CLOUDY model for the O9.5 stars assumed a distance of $100 \pm 15$ pc from the stars to the SNR. The results accounted for a 15 per cent variation in the distance. This was based on the parallax uncertainty of the O9.5 stars (Gaia Collaboration 2020). The SFR and H II regions did not have uncertainties for every region, so the 15 per cent variation was used for consistency (L. Anderson et al. 2014).

The model for the two groups of H II regions assumed a distance of $40 \pm 6$ pc from the groups to the SNR. The UV emission from the H II and star-forming regions was assumed to be produced by stars embedded within the gas. These were assumed to be either one O9.5 star or three B0 stars for each region. (D. Osterbrock & G. Ferland 2006)

In addition to the variation in the distance to the SNR, the density of hydrogen atoms in the ISM clouds was also changed from $1\,\mathrm{cm^{-3}}$ to $100\,\mathrm{cm^{-3}}$ for each run of the model. Although Table 2 lists several parameters that were varied together, we find that the hydrogen density is the dominant controlling parameter for the predicted [C II] emissivity.

The CR ionization rate model is used to examine the predicted impact of CRs on [C II] emission. The CR ionization rate at the SNR needs to be estimated so it can be used in the CLOUDY model. F. Schuppan et al. (2012) used the proton spectra at an SNR to predict the CR ionization rate. Since there is obviously no in-situ measurement of the proton spectrum at the source, they used the GeV $\gamma$-ray emission inferred from hadronic interactions to derive the proton spectrum at an SNR. The details regarding energy loss, particle acceleration, and the normalization of the proton spectrum are described in F. Schuppan et al. (2012). Equations (17)–(19) in F. Schuppan et al. (2012) show the functional form of the proton spectra and their application to the CR ionization rate. F. Schuppan et al. (2012) suggest three possible values for the lower break energy ($E_{lb}$) of the proton spectrum: 0.03 GeV, 0.1 GeV, and 1 GeV. These $E_{lb}$ values account for the variation in the shape of proton spectrum below 1 GeV. This variation is due to the lack of direct observational data inferring the proton spectrum below 1 GeV. The presence of low-energy CRs would lead to a lower $E_{lb}$, as these low-energy CRs are accelerated over the lifetime of the SNR the $E_{lb}$ will increase. Since RX J1713 is a young SNR, we assumed the low-energy CRs were still trapped within the shock, which could suggest $E_{lb}$ values between 0.03 and 0.1 GeV are plausible. Based on the F. Schuppan et al. (2012) model, this suggests a CR ionization rate $\eta_{CR} \sim 10^{-14}\,\mathrm{s^{-1}}$ - $10^{-15}\,\mathrm{s^{-1}}$.

CLOUDY uses $10^{-16}\,\mathrm{s^{-1}}$ as the default Galactic CR ionization rate. We had a maximum $\eta_{CR}$ from F. Schuppan et al. (2012) and a minimum $\eta_{CR}$ from CLOUDY. Therefore, the CLOUDY model tested three CR ionization rates, $\eta_{CR} = 10^{-14}\,\mathrm{s^{-1}}$, $10^{-15}\,\mathrm{s^{-1}}$, and $10^{-16}\,\mathrm{s^{-1}}$.

The output from the CLOUDY models is shown in Fig. 13 and the list of parameters used for the models is shown in Table 2. Of the parameters in Table 2, the variations in the source distances did not have a significant effect on the [C II] emissivity ($\lesssim$ half an order of magnitude). The systematic variation in the [C II] emissivity was driven almost entirely by changes in the assumed hydrogen density, which is therefore the critical parameter; the CR ionization rate sets the overall normalization of the CR-only prediction. Figs H1 and H2 show how the [C II] emissivity from the O9.5 stars, H II regions and star-forming region contribute to the total [C II] emissivity in Fig. 13. The figures represent the upper (Fig. H2) and lower (Fig. H1) limit of the UV emission from CLOUDY. The [C II] emissivity from UV emission in Fig. 13 is the aggregated effect of 15 UV sources. Since the CLOUDY model is 1-dimensional, the contribution of the H II regions and SFRs to the [C II] emissivity is added along the depth of the cloud. At low $R_c$, the [C II] emissivity from UV sources is the total effect of the 2 O9.5 stars, the H II regions and the star-forming region. As $R_c$ increases, the UV photons from the stars lose energy first, which reduces their contribution to the [C II] emissivity. Towards the outer edge of the ISM gas ($\sim 10^{20}$ cm) the [C II] emissivity from UV emission drops to $\sim 10^{-24}\,\mathrm{erg\,cm^{-3}\,s^{-1}}$. This is within the prediction of the [C II] emissivity produced by CRs. The UV photons lose energy as they travel through the ISM gas. Therefore, the deeper into the ISM gas the UV photons have to travel, the less effective the UV ionization. This causes the systematic variation of the [C II] emissivity to narrow towards the outer edge of the ISM gas. It also leads to the crisscross of lines starting at $\sim 10^{19}$ cm in Fig. 13. This suggests that factors affecting the systematic variation (primarily the density of hydrogen) do not have an impact on the [C II] emissivity from UV photons at the outer edge of the gas cloud.

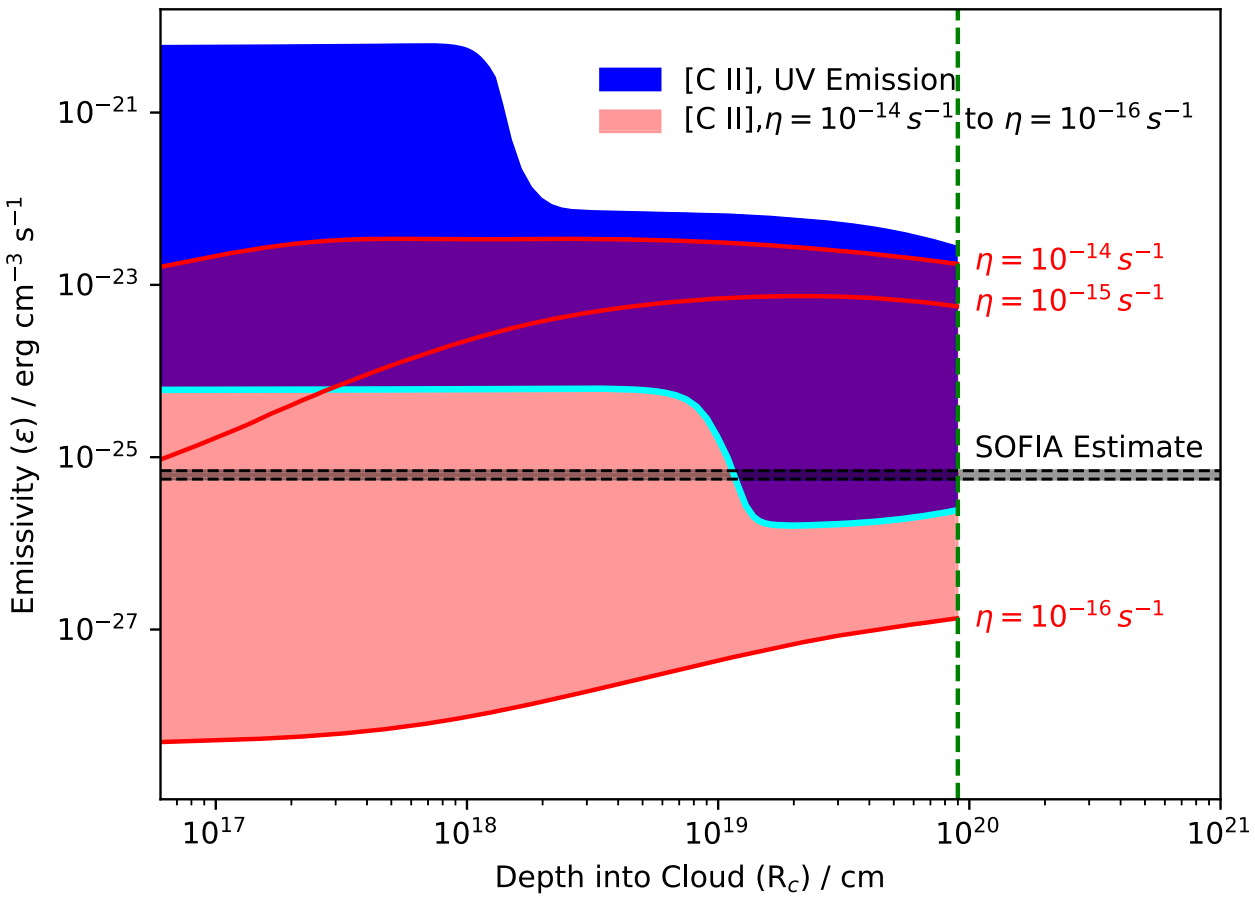

**Figure 13.** [C II] emission predicted by CLOUDY at SNR RX J1713, versus depth of ISM shell $R_c$. The red band is the systematic variation in the [C II] emissivity produced only by CRs. The top of the red band shows the [C II] emissivity assuming a CR ionization rate, $\eta_{CR} = 10^{-14}\,s^{-1}$ and the bottom assumes $\eta_{CR} = 10^{-16}\,s^{-1}$. The line across the middle of the red band assumes $\eta_{CR} = 10^{-15}\,s^{-1}$. The blue band is the systematic variation in the [C II] emissivity produced by UV emission from the UV sources mentioned in Section 6 (O9.5 stars, H II and star-forming regions). The cyan line shows the [C II] emissivity from all UV sources at a hydrogen density of 1 cm$^{-3}$ (Fig. H1). The top of the blue band is the [C II] emissivity from all UV sources at a hydrogen density of 100 cm$^{-3}$ (Fig. H2). The grey band is the [C II] emission calculated based on the integral of the brightness temperature of [C II] from the SOFIA data at region M1 of RX J1713 via equation (1). The width of the grey band accounts for uncertainty in the path length L in equation (1). The $x$-axis shows the depth ($R_c$) into the shell of the cloud, where the outer edge of the cloud (shown by the green dashed line at $\sim 10^{20}$ cm in the figure) is located at SNR RX J1713. The CRs are assumed to enter the cloud from the outer edge, while the UV emission originates from sources at the inner edge of the ISM ring.

In order to compare the predicted [C II] emission from CLOUDY to the [C II] detected by SOFIA, we needed to convert the brightness temperature from SOFIA to an emissivity. The SOFIA [C II] emissivity (erg cm$^{-3}$ s$^{-1}$) is calculated using equation (1).

$$\epsilon = \frac{2kT_b}{L\lambda^2}\Delta\nu\, 4\pi\,, \tag{1}$$

where $T_b$ is the brightness temperature, $\lambda$ is the wavelength of the [C II] transition line, $k$ is Boltzmann's constant, $\Delta\nu$ is the linewidth of the [C II] line emission, and the path length, $L$, is the line-of-sight distance through the emitting [C II] gas, taken as the diameter of the SNR. The grey band for the [C II] emissivity from SOFIA in Fig. 13 results from the uncertainty in the diameter of the SNR which has different estimates in literature, ranging from 16 to 24 pc (F. Aharonian et al. 2006; G. Morlino, E. Amato & P. Blasi 2009; D. Tateishi et al. 2021). The SOFIA [C II] emissivity takes the brightness temperature of [C II] across the M1 region of the SNR. The clumpiness of the gas across the shell may lead to an underestimate of the [C II] emissivity as a filling factor is not accounted for in equation (1).

Fig. 13 suggests the [C II] emissivity at SNR RX J1713 could be attributed solely to the UV emission from nearby sources (mentioned above). While Fig. 13 indicates CR produced within the SNR are capable of producing the [C II] emission at RX J1713, the established UV sources are more likely to be the source of ionization. As a further cross-check, Appendix I compares the UV continuum predicted from our CLOUDY model to an upper limit inferred from *GALEX* (P. Morrissey et al. 2007) observations towards SNR RX J1713. We find that our CLOUDY prediction has a value $\sim$ 0.85 times that of the *GALEX* upper limit.

## 7 DISCUSSION

The I[C II]/N[$H_2$] and I[C II]/N[H I] ratios, shown in Fig. 5, show the [C II] emission generally following the H I while being anti-correlated with the $H_2$. The profiles in Fig. 5 suggest that the level of ionized carbon in the clouds depends on the column density of atomic gas. The gas on the left side of the remnant has a higher atomic fraction as well as a higher I[C II]/N(H I) ratio. This could indicate that the denser gas on the right of the remnant is not being ionized at the same level as the gas on the left. The gas could be ionized by X-rays, UV radiation and low-energy CRs. The X-rays can penetrate deep into the clouds and ionize them effectively (A. Tielens 2005; W. D. Langer & J. L. Pineda 2015). X-rays with energies of 1 keV have a penetration depth $> 10^{22}$ cm$^{-2}$ in diffuse ionized gas and $\sim 10^{21}$ cm$^{-2}$ in neutral gas (W. D. Langer & J. L. Pineda 2015). The observed X-ray emission from *XMM–Newton* (J. S. Hiraga et al. 2005) and Chandra (Y. Uchiyama, F. A. Aharonian & T. Takahashi 2003) show a hydrogen column density ($N_H$) of $\sim$0.7-0.9 $\times 10^{22}$ cm$^{-2}$ across the SNR, which suggests the X-ray emission cannot penetrate the neutral gas at the SNR. We see a stronger I[C II] signal on the left side of the SNR in Fig. E1. This suggests the source of ionization cannot penetrate dense gas, such as the molecular core of gas on the right side (l = 347.1, b = -0.4) in Fig. E1. UV photons and low-energy CRs would ionize the outer layer of gas effectively, but are unable to penetrate deep into the gas (A. Tielens 2005). This effect can be seen for the UV photon sources in Figs H1 and H2. The [C II] emissivity for UV photons begins to fall off at a depth of $\sim 10^{16}$ cm for hydrogen densities of 100 cm$^{-3}$, as opposed to a depth of $\sim 10^{19}$ cm for densities of 1 cm$^{-3}$.

## 8 CONCLUSIONS

The high-resolution I[C II] data from the SOFIA telescope have given us a new look at SNR RX J1713. We have been able to compare the distribution of I[C II] to other atomic and molecular gas data and found that the I[C II] signal seems to vary across the remnant. Our analysis suggests that the I[C II] signal is stronger in atomic clouds, as opposed to molecular clouds. This could indicate the presence of low-energy CRs or UV photons, which might be able to more easily penetrate the less dense atomic gas and ionize it.

The GOT C+ data show that the ratio of I[C II]/I[$^{12}$CO] does not vary significantly across the galaxy for star formation regions, H II regions and supernova remnants. This suggests that the levels of ionization are similar across these regions. However, larger observation areas could reveal more about the distribution of [C II] in these regions, as it has with the SOFIA data.

The CLOUDY model shows an overlap in the predicted [C II] emissivity from CRs and UV photons at SNR RX J1713. This suggests UV photons and CRs could each be responsible for the ionization of carbon at RX J1713, assuming low-energy CRs are trapped within the SNR. However, the predicted UV emission from the observed stars and H II regions is sufficient to produce the [C II] emissivity observed at RX J1713. We need to study other SNRs located in different parts of the galaxy, with potentially different UV emission. Analysis of other SNRs could also help identify the role low-energy CRs play in the production of [C II]. We are currently working with SOFIA data taken from RCW86 and Vela Jr (RX J0852.0−4622). Since the SOFIA data are more sensitive than those provided by other telescopes that probe this line, they might reveal more about the ionization caused by these objects.

## ACKNOWLEDGEMENTS

We thank the anonymous referee for their constructive comments, which improved the clarity and accuracy of this work.

This work is based on observations made with the NASA/DLR Stratospheric Observatory for Infrared Astronomy (SOFIA). SOFIA was jointly operated by the Universities Space Research Association, Inc. (USRA), under NASA contract NNA17BF53C, and the Deutsches SOFIA Institut (DSI) under DLR contract 50 OK 0901 to the University of Stuttgart.

The Mopra radio telescope is part of the Australia Telescope National Facility, which is funded by the Australian Government for operation as a National Facility managed by CSIRO. Operations support for the Mopra CO survey was provided by the University of New South Wales, the University of Adelaide, and the National Astronomical Observatory of Japan.

This research made use of data from the Southern Galactic Plane Survey, conducted with the Australia Telescope Compact Array and the Parkes radio telescope, which are part of the Australia Telescope National Facility, funded by the Australian Government and managed by CSIRO. We acknowledge the Wiradjuri people as the Traditional Owners of the Parkes Observatory site.

This work also made use of data from H.E.S.S., *XMM–Newton*, the Wisconsin H-Alpha Mapper (WHAM) survey, the *Herschel* GOT C+ survey and *GALEX*, and of the CLOUDY photoionization code (G. J. Ferland et al. 2017).

ART acknowledges support from a University of Adelaide scholarship.

## DATA AVAILABILITY

The observational data used in this study – including SOFIA [C II] line maps, Mopra CO molecular gas data, SGPS H I atomic gas data, and high-energy maps from H.E.S.S. and *XMM–Newton* – are publicly available from their respective archives. SOFIA data can be accessed via the SOFIA Science Center, Mopra CO data via the Mopra Data Archive, SGPS H I data via the Australian Telescope Online Archive, and H.E.S.S./*XMM–Newton* maps via the H.E.S.S. Data Release Portal and *XMM–Newton* Science Archive. Analysis scripts and results are available from the corresponding author upon reasonable request.

## APPENDIX A: SOURCES OF ENERGETIC RADIATION

Table A1 summarizes the photon and cosmic-ray populations referred to throughout this work, together with their approximate energy ranges, the observational tracers used to identify them, and their relevance to the ionization of carbon in the ISM.

**Table A1.** Summary of the energetic photons and cosmic-ray populations referred to in this work, their approximate energy ranges, how they are traced, and their relevance to ionization of the ISM.

| Category | Approx. energy | Traced by | Relevance to this work |
|---|---|---|---|
| Ionizing UV (EUV) | $> 13.6$ eV | H $\alpha$ | Ionizes hydrogen, producing H II regions and H $\alpha$ emission; used here to locate candidate UV sources (Fig. 11). Absorbed at the ionization front near the cloud edge and does not penetrate the molecular gas. |
| Far-UV (FUV) | 6–13.6 eV | [C II]; PDR modelling | Below the hydrogen ionization threshold, so it does not ionize hydrogen (no H $\alpha$), but photoionizes carbon (11.3 eV) and other heavy elements within photodissociation regions, producing $C^+$ at low ionization fraction ($\sim 10^{-4}$ to $10^{-6}$). Remains the dominant source of $C^+$ to large depths; the transition to CR-dominated ionization is density-dependent, reaching $A_V \sim 10$ in dense gas. |
| X-rays | $\sim$ 0.1–10 keV | X-ray continuum emission | Penetrate deep into clouds and ionize carbon throughout, but are attenuated by the high column densities ($N_H \sim 10^{22}$ cm$^{-2}$). |
| Low-energy/sub-GeV CRs | $\lesssim 1$ GeV | [C II] ionization (this work); no direct $\gamma$-ray tracer | ionize the outer layers of clouds but lack the energy to penetrate dense cores; their origin cannot be traced directly, motivating [C II] as an indirect tracer. |
| Multi-GeV CRs | $\gtrsim$ 1–10 GeV | GeV $\gamma$-ray emission | Produce $\gamma$-rays via interaction with ISM gas. |
| GeV–TeV $\gamma$-rays | $\sim$ 0.1 GeV–100 TeV | GeV–TeV $\gamma$-ray emission | Trace multi-GeV CRs. GeV resolution ($\gtrsim 1^\circ$) is too coarse to pinpoint sub-GeV CR origins, whereas TeV resolution allows CR acceleration sites and regions with potentially significant CR populations to be identified. |

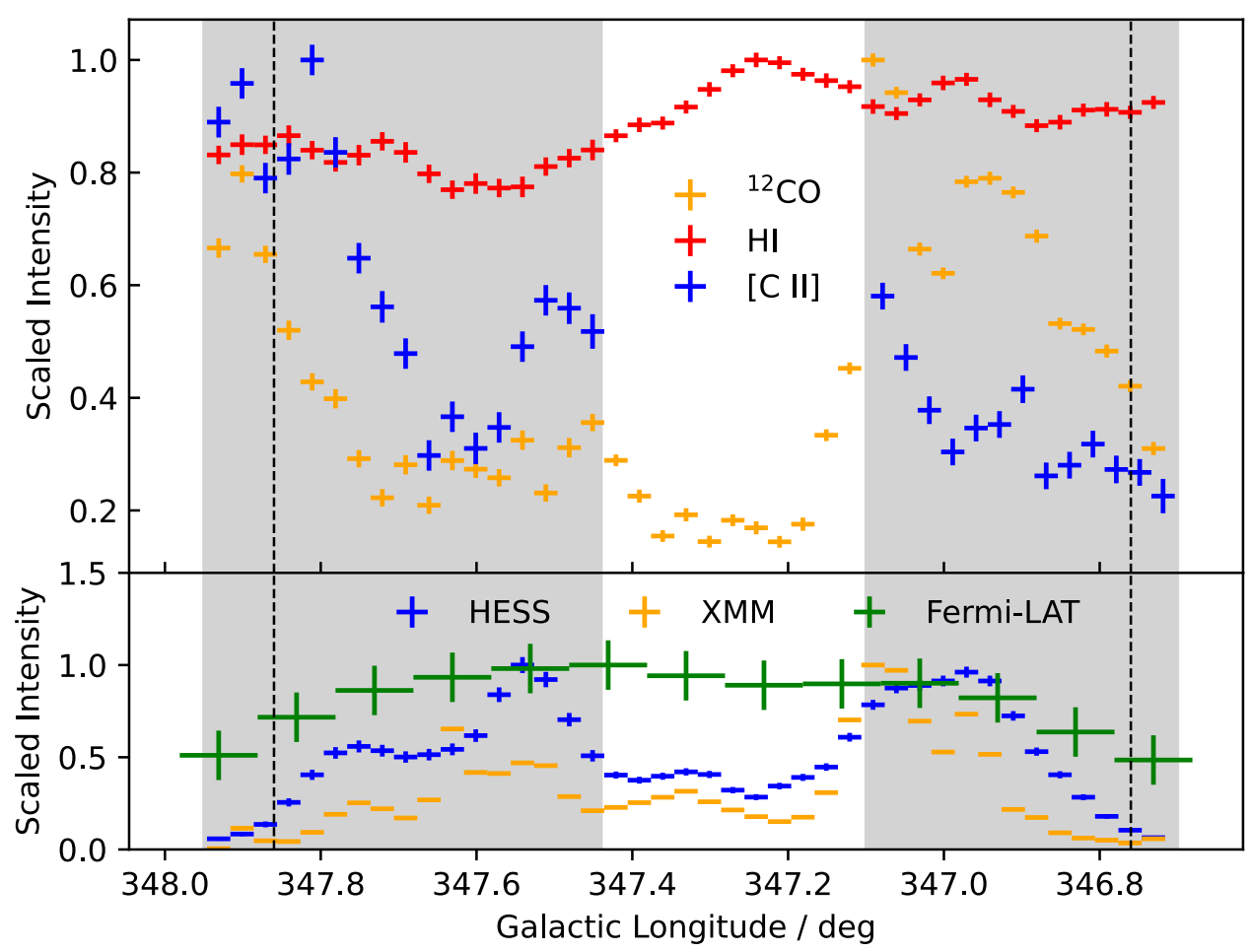

**Figure B1.** *Upper panel:* Mopra I[$^{12}$CO(J = 1–0)], SOFIA I[C II] and SGPS N(H I) longitude profiles across the M1 region (shaded area) of SNR RX J1713, integrated over a velocity range of $-25$ km s$^{-1}$ to 0 km s$^{-1}$. The profiles are scaled to their maximum value. The dashed black lines indicate the boundaries of the SNR, based on the TeV gamma-ray contours shown in Fig. 1. *Lower panel:* The profile of H.E.S.S. $\gamma$-rays (> 2 TeV), *XMM–Newton* X-rays, and *Fermi*-LAT $\gamma$-rays (500 MeV–300 GeV), scaled to their maximum values, across the remnant over the same regions as the SOFIA [C II] M1 mapping.

## APPENDIX B: PROFILES OF GAS OBSERVATIONS ACROSS REGION M1

Fig. B1 shows the longitude profiles of I[C II], I[$^{12}$CO(J = 1–0)] and N(H I) across the M1 region, integrated over $-25$ to 0 km s$^{-1}$ and scaled to their maximum values, alongside the corresponding H.E.S.S., XMM–*Newton* and *Fermi*-LAT profiles. The dashed lines mark the boundaries of the remnant as defined by the TeV $\gamma$-ray contours in Fig. 1.

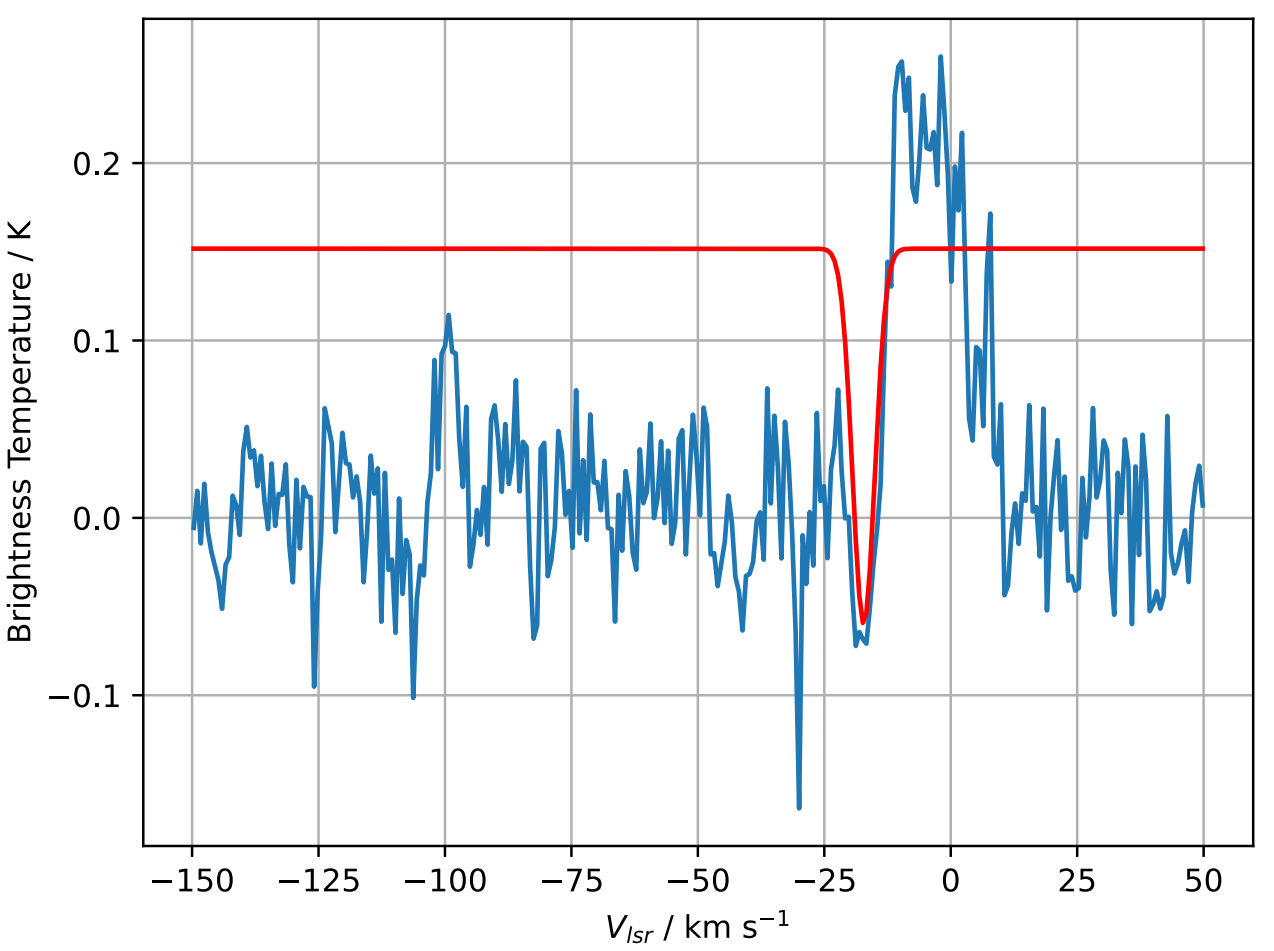

**Figure C1.** Gaussian correction of the dip in the M1 region of the SOFIA [C II] spectra. The offset of ~0.15 K shown in the figure is subtracted when applying the correction to the spectra.

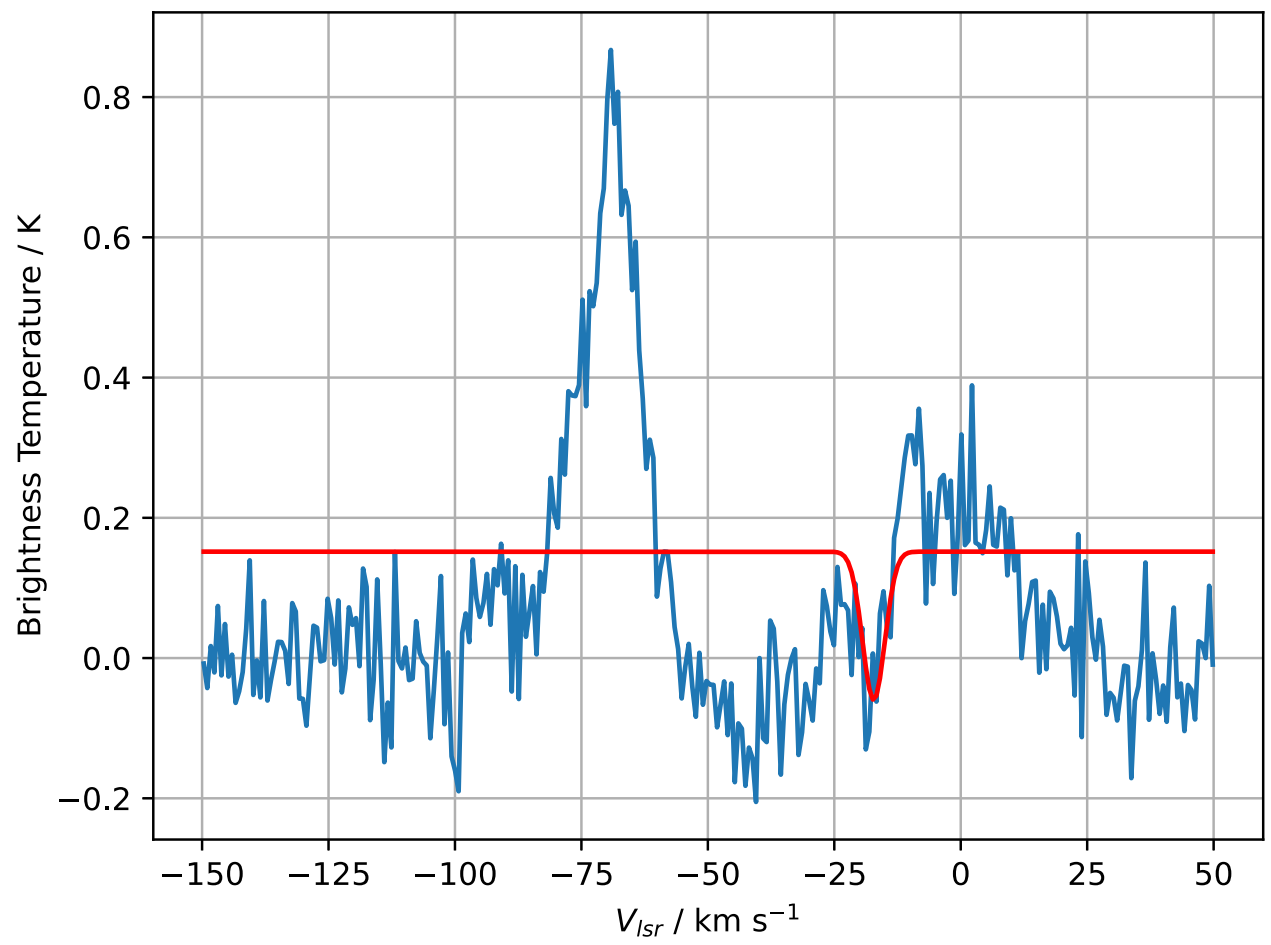

**Figure C2.** Gaussian correction of the dip in the M3 region of the SOFIA [C II] spectra. The offset of ~0.15 K shown in the figure is subtracted when applying the correction to the spectra.

## APPENDIX C: GAUSSIAN MODELS FOR M1 AND M3 CORRECTIONS

Figs C1 and C2 show the Gaussian models used to correct the negative dip present at $\sim -20$ km s$^{-1}$ in the SOFIA [C II] spectra, which arises from [C II] emission in the OFF position. The models were fitted to the M1 and M3 spectra, respectively; the constant offset of $\sim$ 0.15 K shown in each figure is subtracted before the correction is applied to the data.

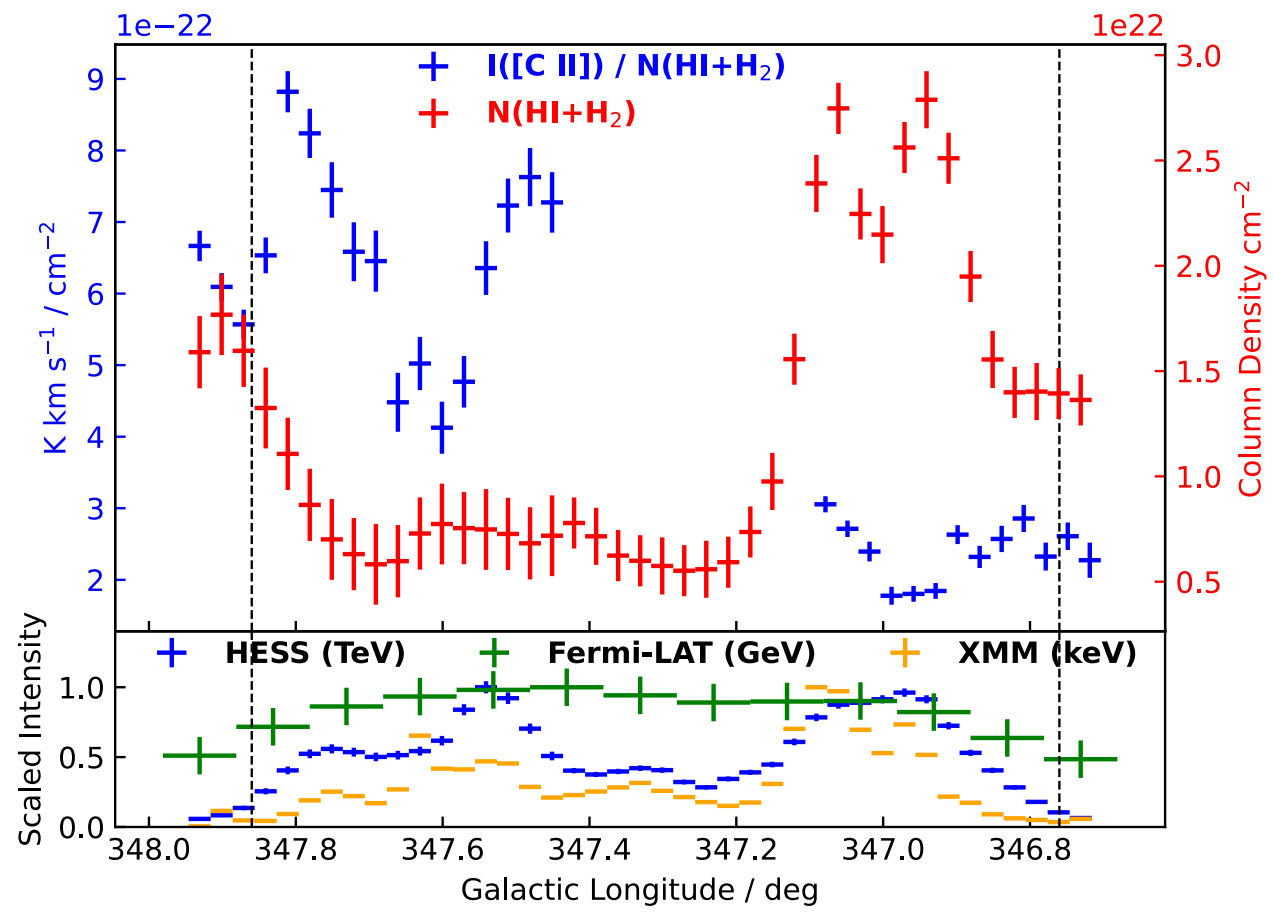

**Figure D1.** Longitude profiles of the ratio of integrated I[C II] to the total gas column density across region M1 of the SNR RX J1713 (*blue*), and the variation in the total column densities (red). The scales and labels in red refer to the column density of N(H I+$H_2$), while the blue label refers to the ratio of I[C II] to the total column density. The black dashed lines show the outer edge of SNR RX J1713, based on the *XMM–Newton* X-ray image. The spectra were integrated across the velocity range of $-20\,\mathrm{km\,s^{-1}}$ to $0\,\mathrm{km\,s^{-1}}$. *Bottom panel*: The profile of H.E.S.S. $\gamma$-rays ($> 2$ TeV), *XMM–Newton* X-rays, and *Fermi*-LAT $\gamma$-rays (500 MeV–300 GeV), scaled to their maximum values, across the remnant over the same regions as the SOFIA [C II] M1 mapping.

## APPENDIX D: LONGITUDE PROFILE OF I[C II]/N(H I+$H_2$)

Fig. D1 shows the ratio of the integrated I[C II] to the total gas column density, N(H I+$H_2$), across region M1, together with the variation in the total column density itself. This complements Fig. 5, which separates the atomic and molecular contributions.

## APPENDIX E: VELOCITY CHANNEL MAP OF [C II]

Figs E1–E3 show velocity channel maps of the SOFIA I[C II] emission for regions M1, M2, and M3. Each panel is integrated over a $5\,\mathrm{km\,s^{-1}}$ interval, with successive panels offset by $2.5\,\mathrm{km\,s^{-1}}$, spanning $-25$ to $0\,\mathrm{km\,s^{-1}}$; the final panel of each figure shows the emission integrated over the full range.

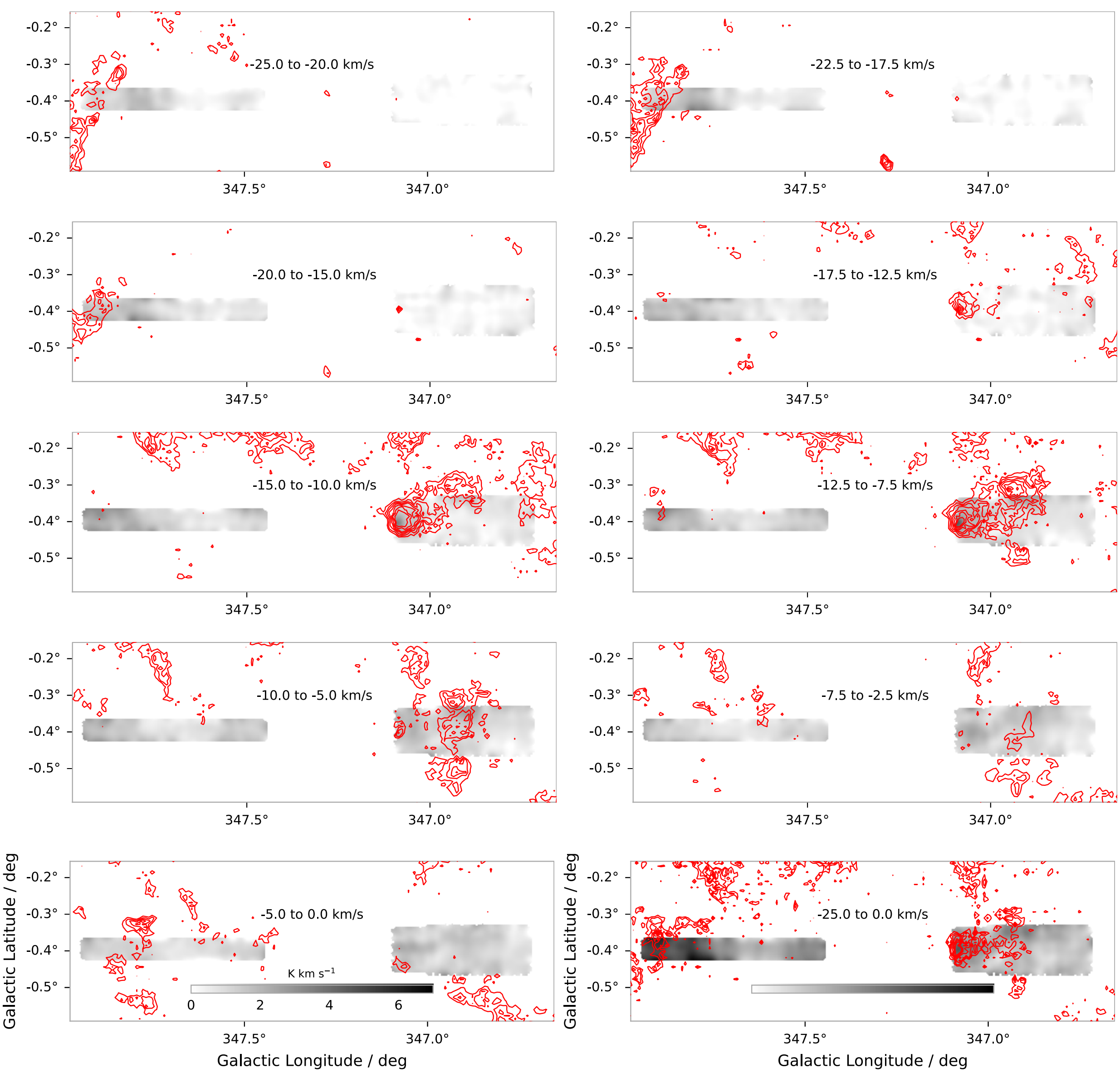

**Figure E1.** Velocity channel maps of the SOFIA I [C II] line emission from the M1 region. The red contours (10–35 K) show the Mopra $^{12}$CO(J = 1–0) emission for the same velocity range. The brightness temperature of 10 K corresponds to $2T_{RMS}$. The last panel is the average SOFIA I [C II] line emission from the M1 region across the entire velocity range.

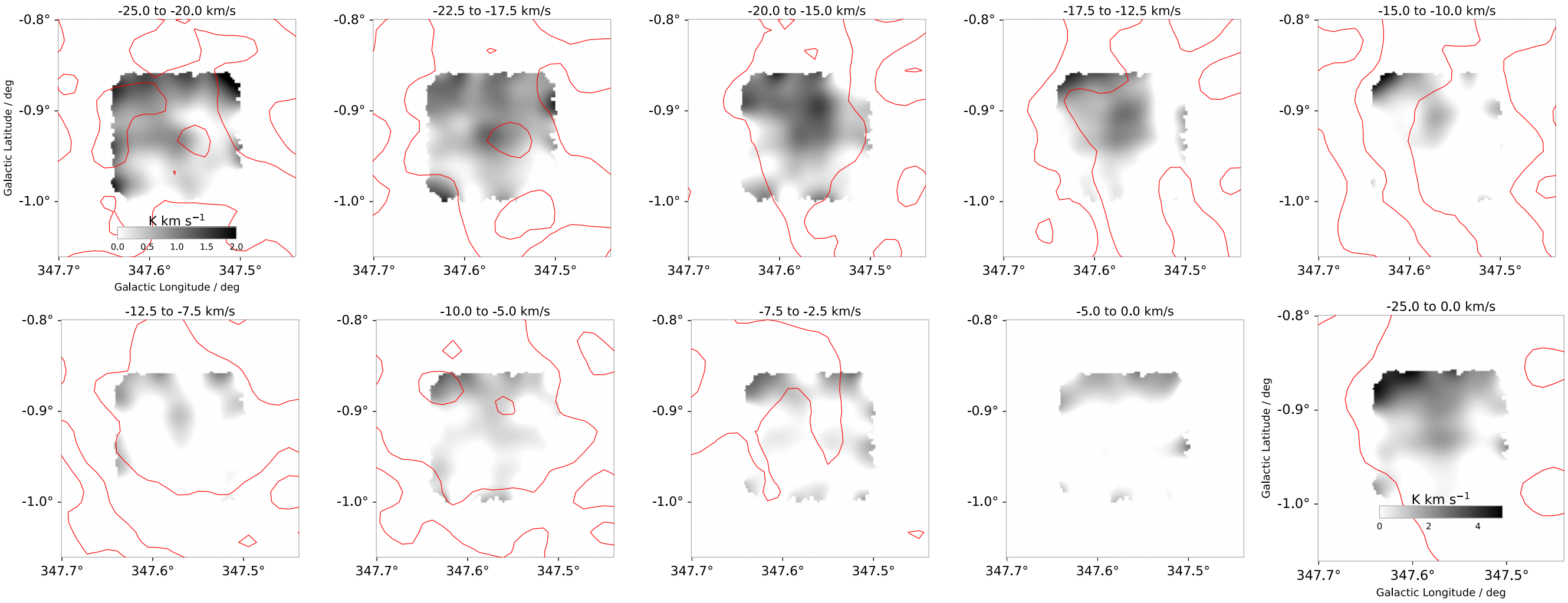

**Figure E2.** The velocity channel maps of the SOFIA I [C II] line emission from the M2 region. The red contours ($5\times10^{21}$ cm$^{-2}$ to $6.4\times10^{21}$ cm$^{-2}$) show the H I SGPS emission for $-20$ to 0 km s$^{-1}$. The column density of $5\times10^{21}$ cm$^{-2}$ corresponds to 2T$_{\rm RMS}$.

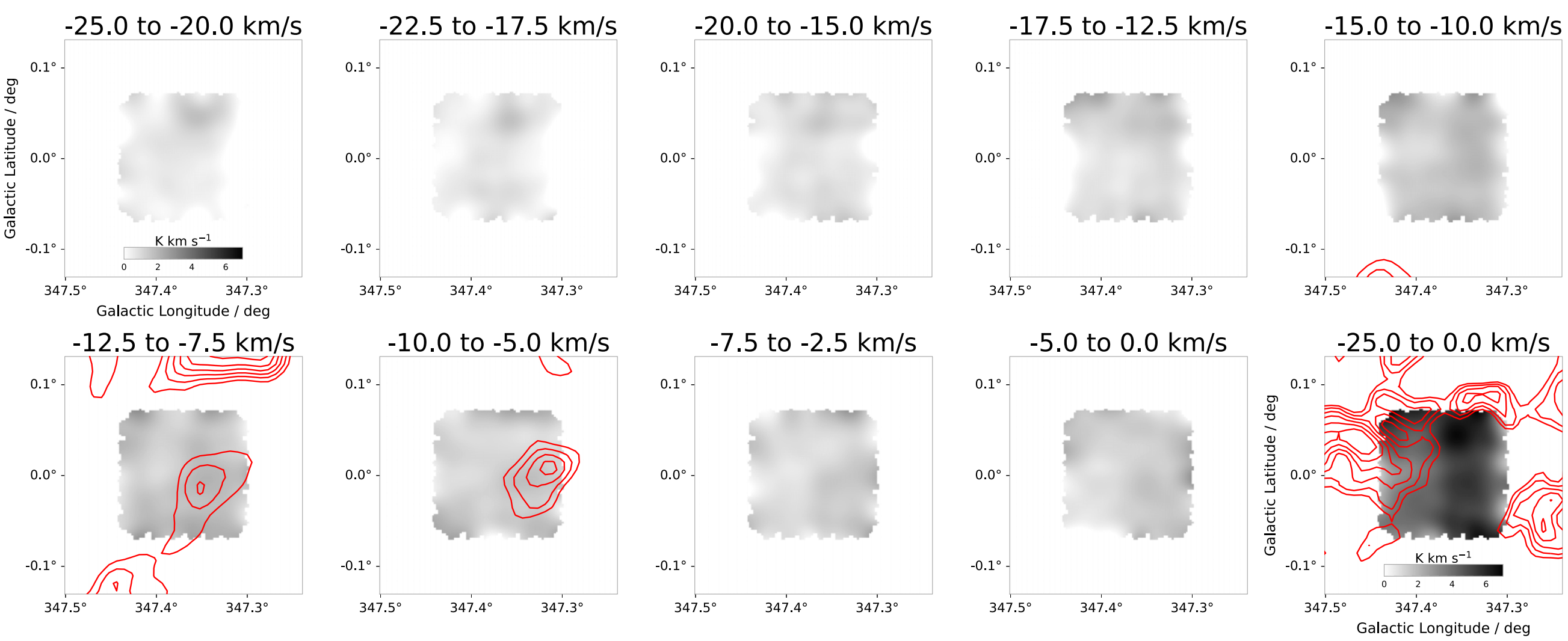

**Figure E3.** The velocity channel maps of the SOFIA I [C II] line emission from the M3 region. The red contours (8–17 K) show the Mopra $^{12}$CO(J = 1–0) emission for the same velocity range. The brightness temperature of 8 K corresponds to 2T$_{\rm RMS}$.

## APPENDIX F: SPECTRA PLOTS FOR M1 SUB-REGIONS

The figures in this section show the spectra across the velocity range of $-150$ km s$^{-1}$ to 50 km s$^{-1}$ for each SOFIA M1 region.

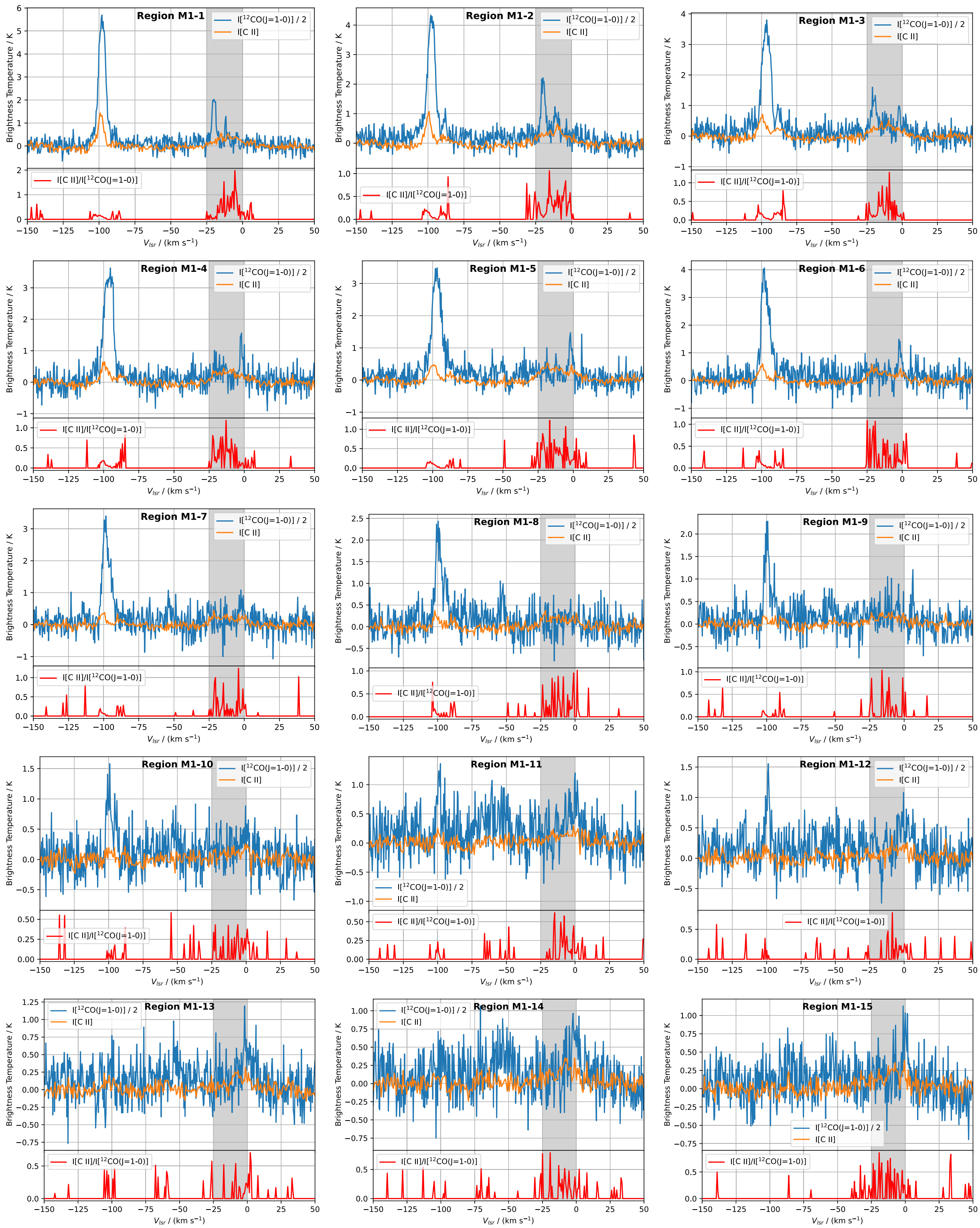

**Figure F1.** Mopra I[$^{12}$CO(J = 1–0)], SOFIA I([C II]) and their ratio from M1 regions 1–15. The C$^+$ and $^{12}$CO spectra are averaged over the corresponding M1 regions, and $^{12}$CO is divided by 2 for better comparison with C$^+$. The shaded band identifies the velocity range within which the gas clouds associated with the SNR RX J1713.7−3946 are located and is our velocity range of interest. The spectra have been smoothed using a $\sim 2$ km s$^{-1}$ wide sliding window.

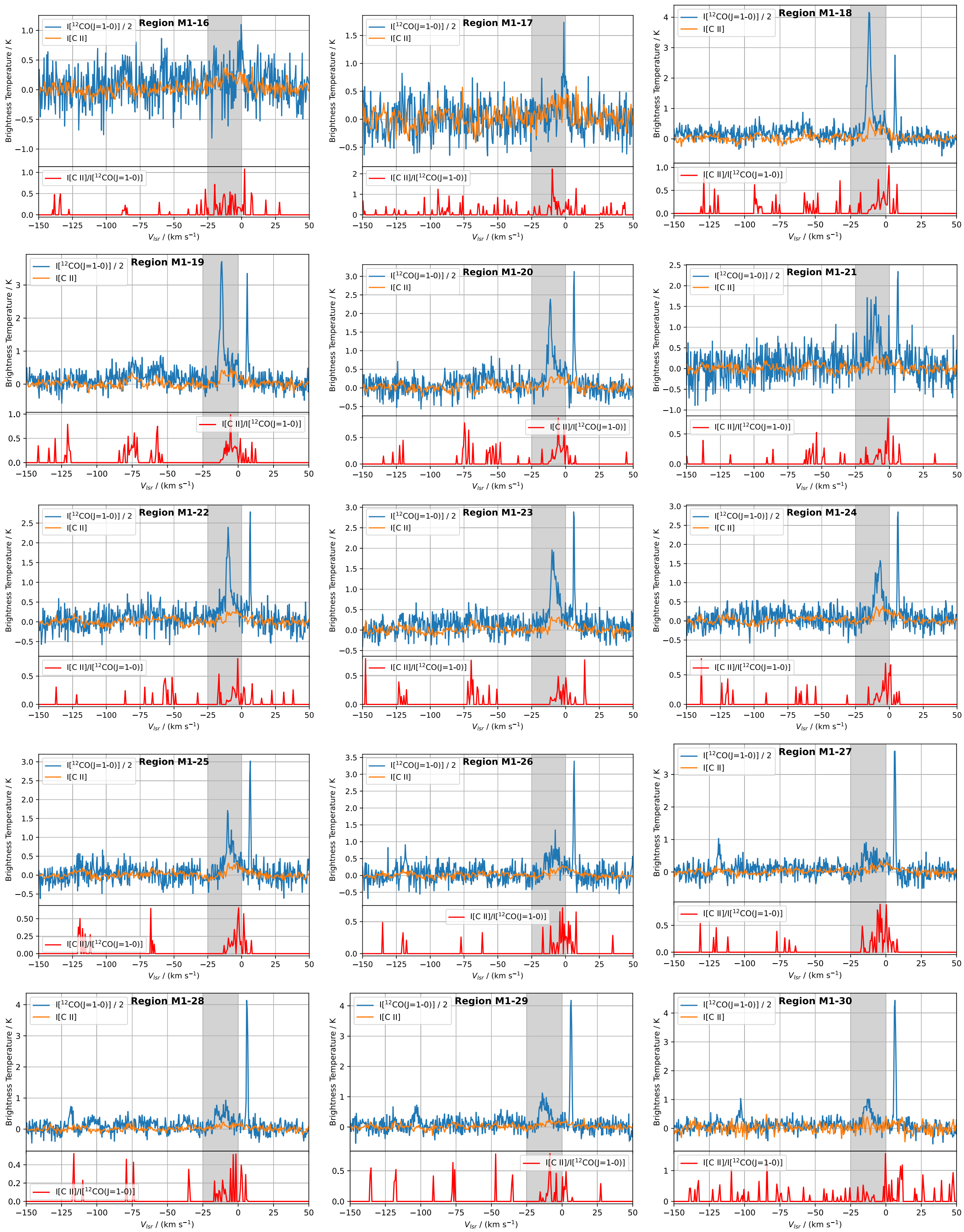

**Figure F2.** Mopra I[$^{12}$CO(J = 1–0)], SOFIA I([C II]) and their ratio from M1 regions 16–30. The $C^+$ and $^{12}$CO spectra are averaged over the corresponding M1 regions, and $^{12}$CO is divided by 2 for better comparison with $C^+$. The shaded band identifies the velocity range within which the gas clouds associated with the SNR RX J1713.7−3946 are located and is our velocity range of interest. The spectra have been smoothed using a $\sim$2 km s$^{-1}$ wide sliding window.

**Table G1.** Median values of the I[C II]/I[$^{12}$CO(J = 1–0)] distributions for GOT C+ and SOFIA.

| GOT C+ | SOFIA (−25 to 0 km s$^{-1}$) | SOFIA (−100 to −75 km s$^{-1}$) |
|---|---|---|
| 0.14 ± 0.01 | 0.49 ± 0.09 | 0.35 ± 0.08 |

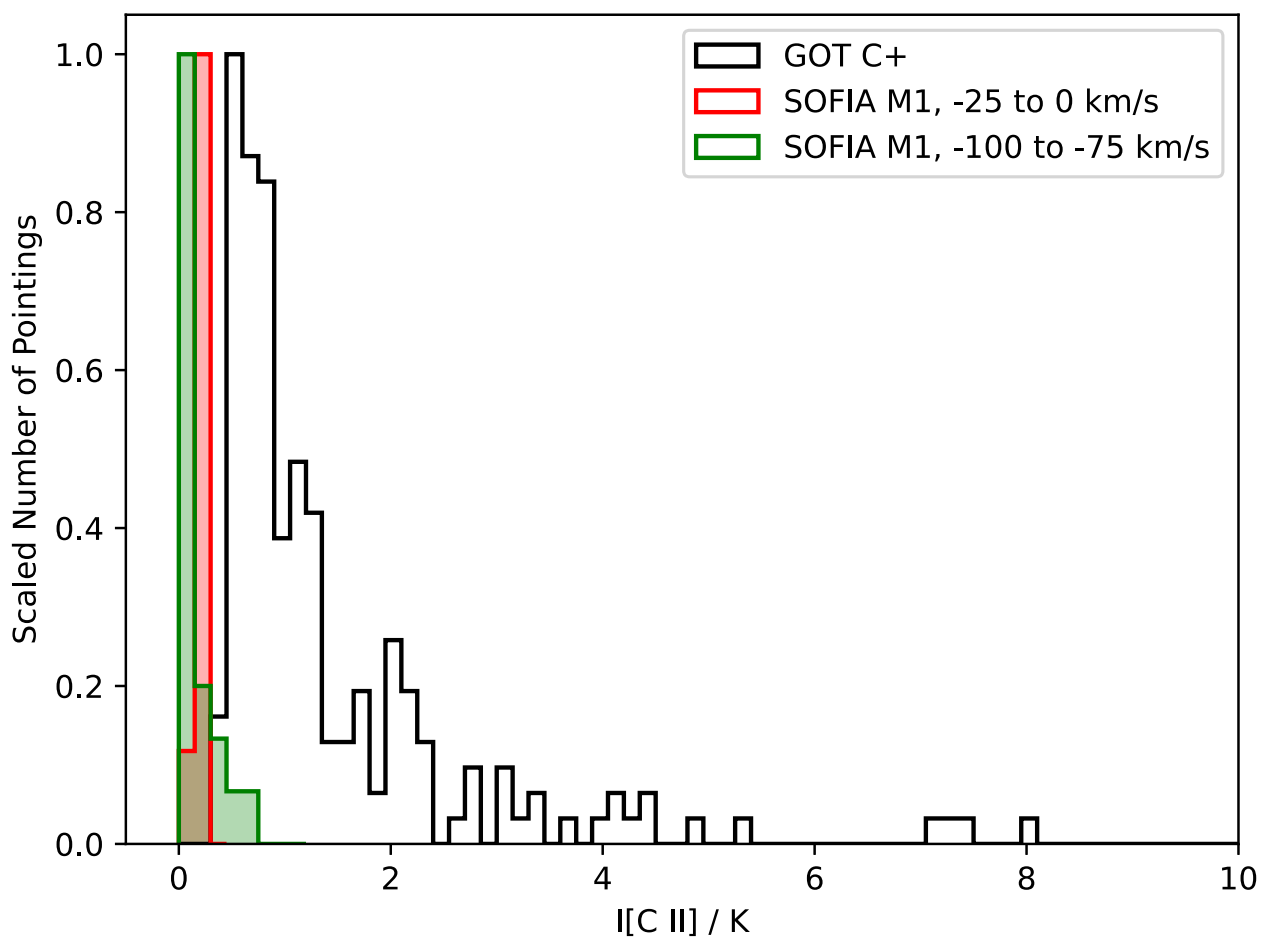

**Figure G1.** Distribution of I[C II] (average of the peak and its two adjacent channels) from the GOT C+ survey and our M1 SOFIA data. The distributions include RX J1713 (−25 to 0 km s$^{-1}$; *red*), the molecular gas cloud behind RX J1713 (−100 to −75 km s$^{-1}$; *green*, and the GOT C+ survey *(black)*. The histograms are scaled to the maximum value in each distribution.

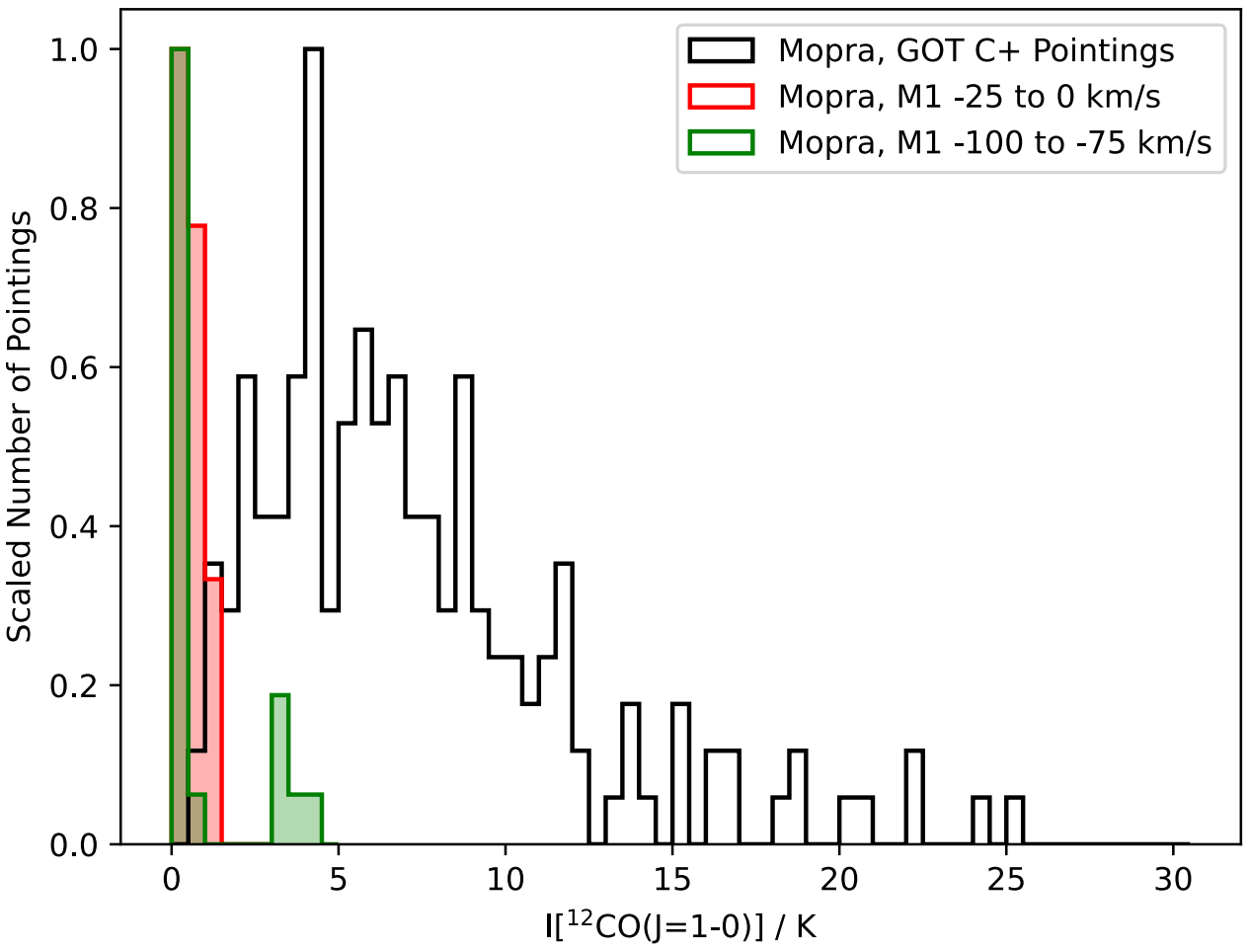

**Figure G2.** Distribution of I[$^{12}$CO(J = 1–0)] (average of the peak and its neighbouring channels) from Mopra data from GOT C+ pointing and the M1 region. The distributions include RX J1713 (−25 to 0 km s$^{-1}$; *red*), the molecular gas cloud behind RX J1713 (−100 to −75 km s$^{-1}$; *green*), and the GOT C+ survey (*black)*. The histograms are scaled to the maximum value in each distribution.

## APPENDIX G: GOT C+ HISTOGRAMS

We extracted the GOT C+ [C II], Mopra $^{12}$CO, and SGPS H I emission averaged around their spectral peak values. The line intensities (K) in the histograms (Figs G1, G2, and G4) are the average of the peak and its two adjacent velocity channels, to account for the width of the peak. We note that there are some pointings with more than one peak. The ratio of the [C II] to $^{12}$CO and H I spectral peaks found in each GOT C+ pointing are shown in Figs G3 and G5. Table G1 shows the median values of the distributions.

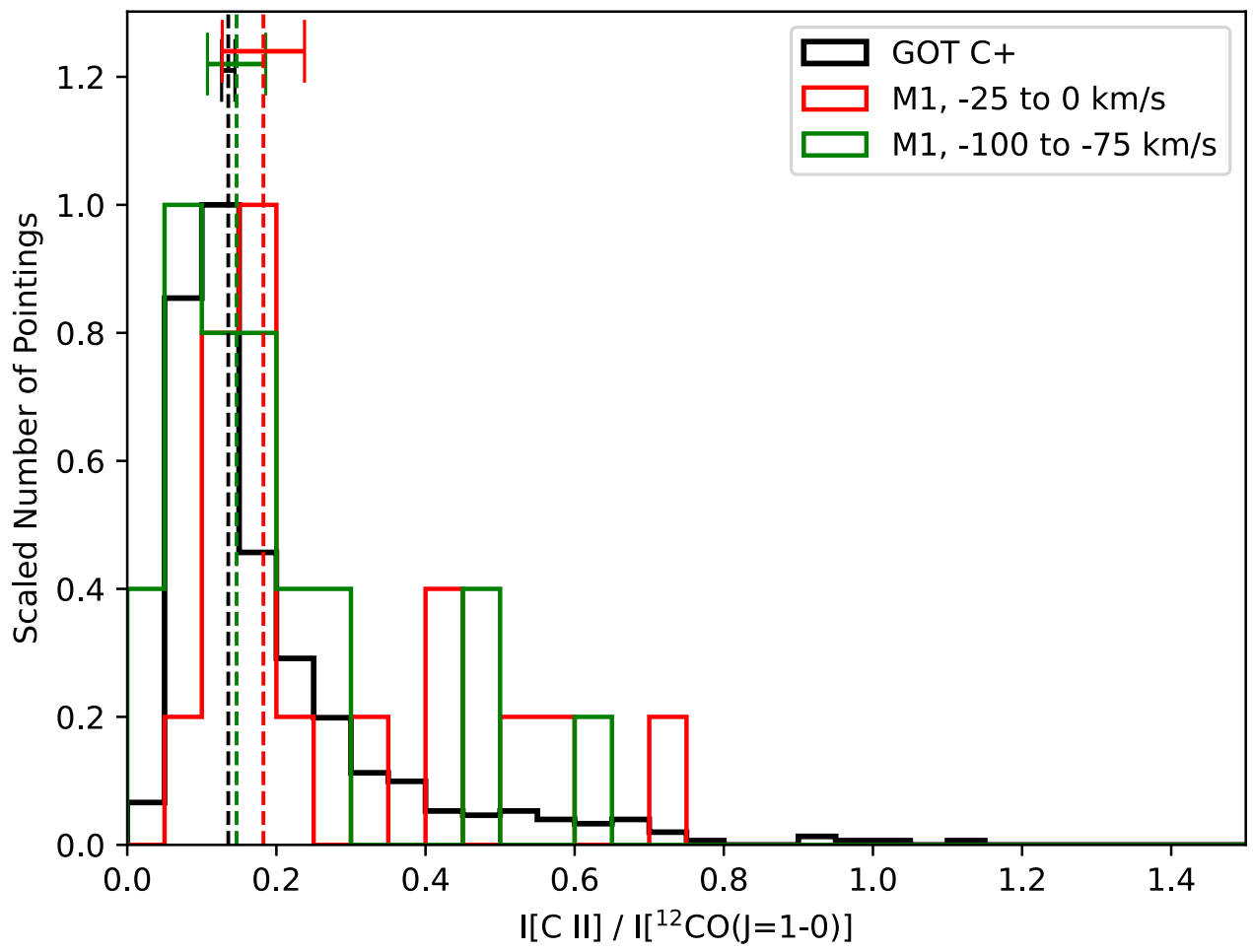

**Figure G3.** Distribution of scaled *Herschel* I[C II]/I[$^{12}$CO(J = 1–0)] ratios taken from GOT C+ and Mopra $^{12}$CO for the M1 region (I[C II] and I[$^{12}$CO] distributions shown in Figs G1 and G2). To qualify as significant detection, an emission feature needed to be 3$\sigma$ above noise. The black histograms are the ratios taken from the GOT C+ survey. The −25 km s$^{-1}$ to 0 km s$^{-1}$ range includes SNR RX J1713 (red), and the −100 km s$^{-1}$ to −75 km s$^{-1}$ range is believed to be a background molecular gas cloud (green). The vertical dashed lines indicate the median of the distribution with the corresponding colour, and the error bars indicate the standard error in the median. The histograms are scaled to the maximum value in each distribution.

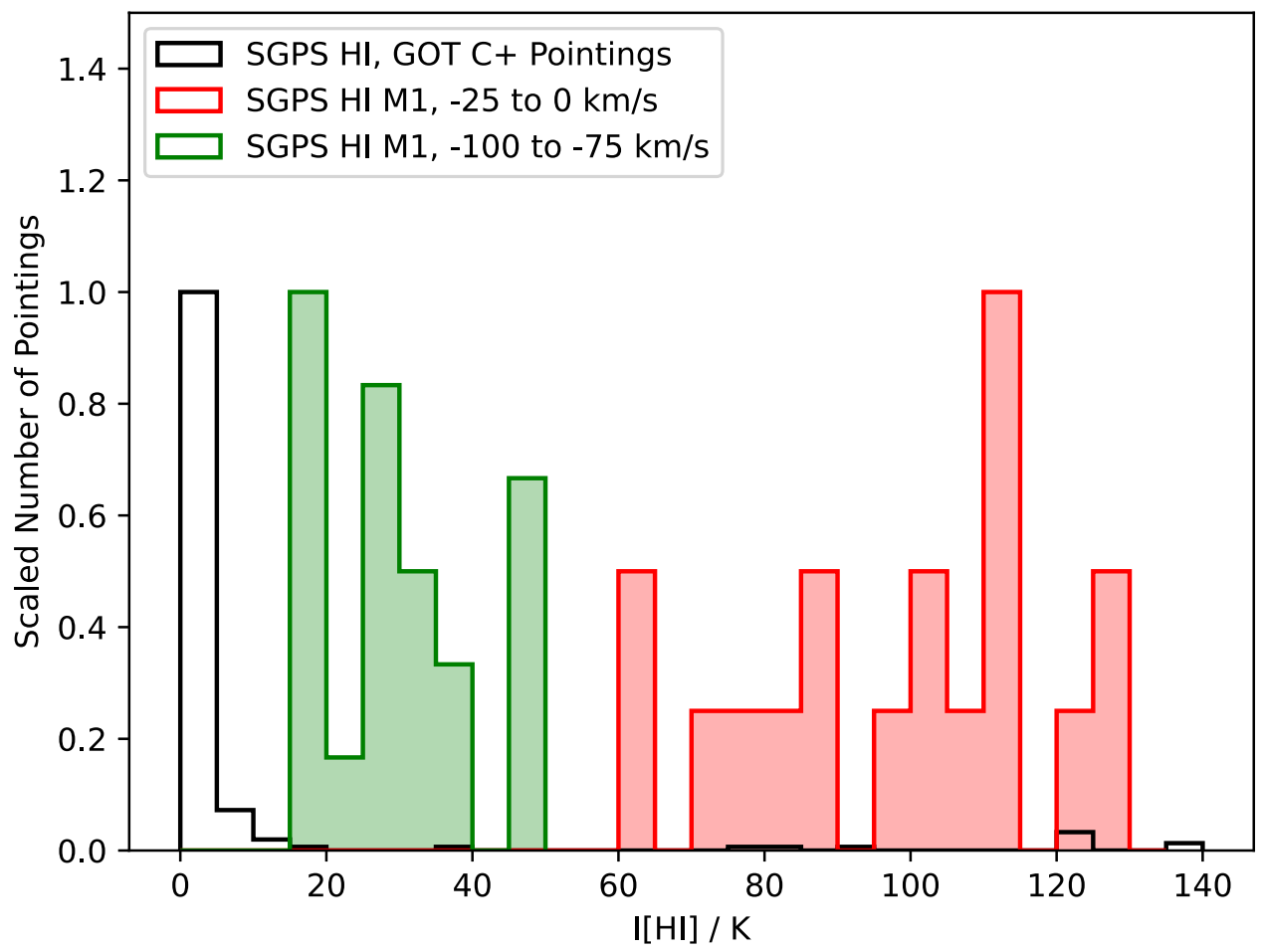

**Figure G4.** The distribution of I[H I] from GOT C+ and from the RX J1713 M1 region. The distributions include RX J1713 (*red*), the molecular gas cloud behind RX J1713 (*green*), and the GOT C+ survey (*black*). The histograms are scaled to the maximum value in each distribution.

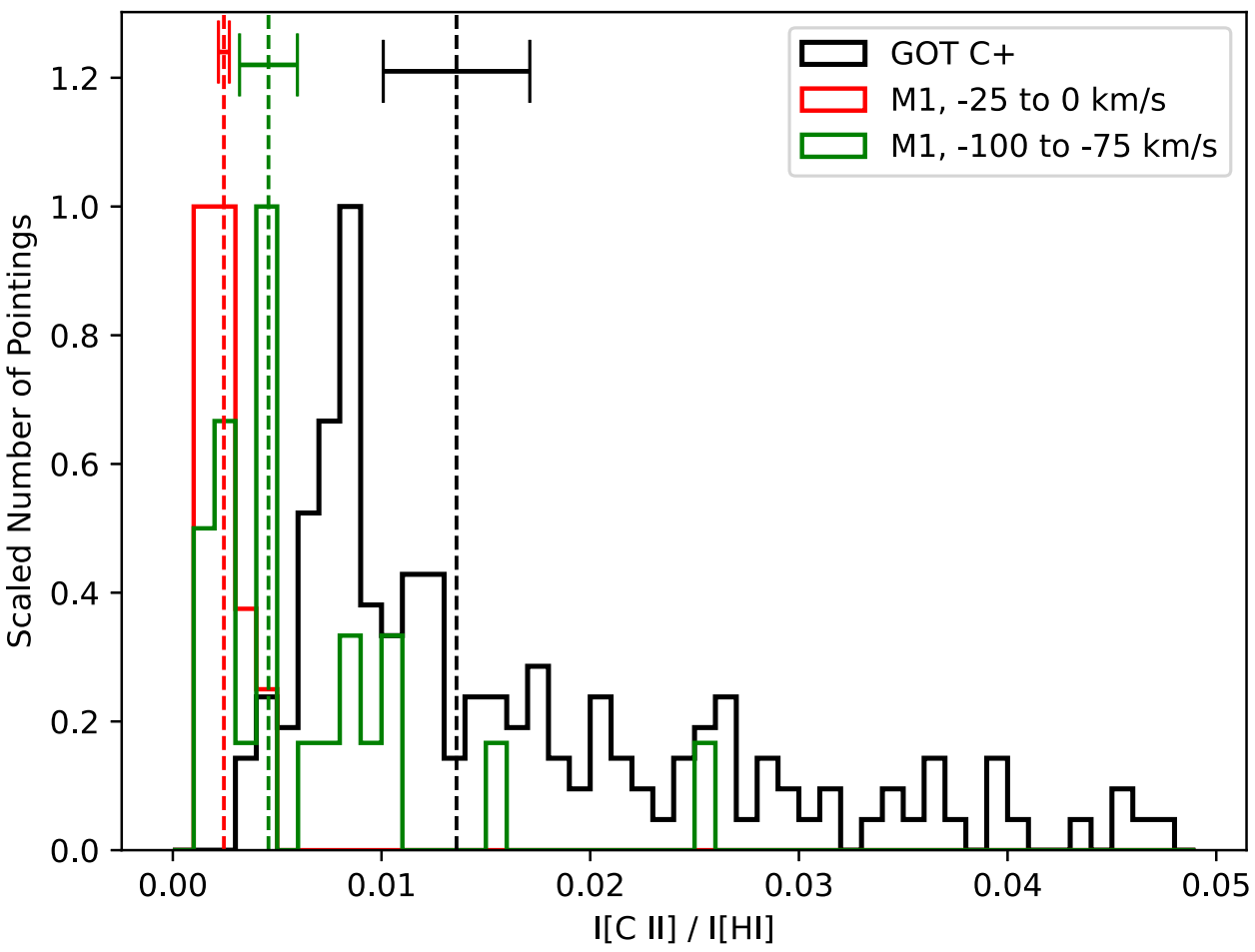

**Figure G5.** Distribution of I[C II]/I[H I] ratios taken from GOT C+ and SGPS (I[C II] and I[H I] distributions shown in Figs G1 and G4). The black columns are the ratios taken from the GOT C+ survey. The $-25\,\mathrm{km\,s^{-1}}$ to $0\,\mathrm{km\,s^{-1}}$ range includes SNR RX J1713 (red) and the $-100\,\mathrm{km\,s^{-1}}$ to $-75\,\mathrm{km\,s^{-1}}$ range (green) includes the background molecular cloud. The vertical dashed lines indicate the median of the distribution with the corresponding colour, and the error bars indicate the standard error in the median. The histograms are scaled to the maximum value in each distribution.

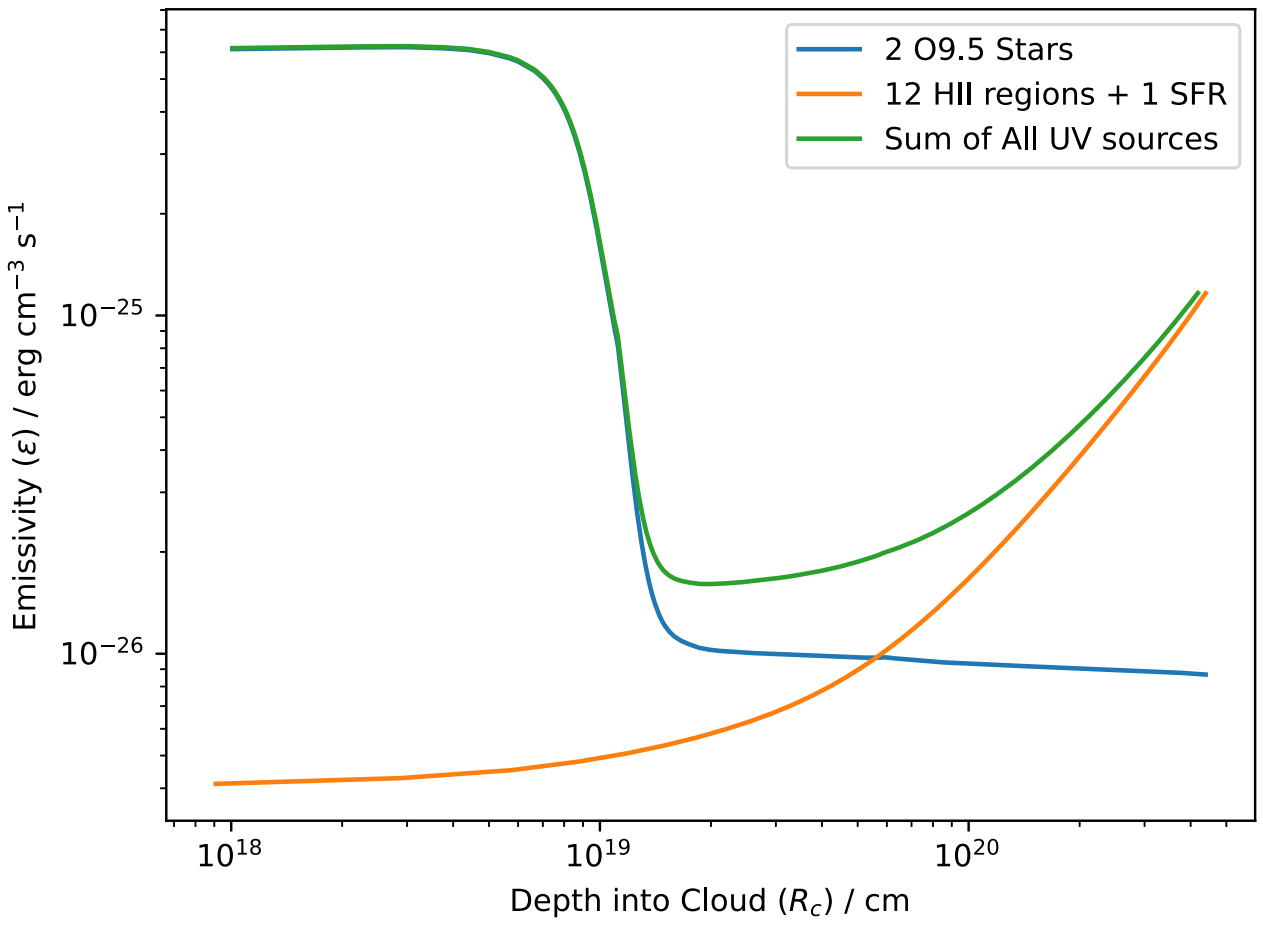

**Figure H1.** [C II] emission predicted by CLOUDY at SNR RX J1713, versus depth of ISM shell $R_c$ with $n_H = 1\,\mathrm{cm^{-3}}$. The blue line shows the [C II] emissivity from 2 O9.5 UV stars (100 pc away from the SNR), while the orange line shows the [C II] emissivity from the H II and star-forming regions (40 pc away). The total [C II] emissivity from all UV sources is shown by the green line. The *x*-axis shows the depth ($R_c$) into the shell of the cloud, where the outer edge of the cloud (shown by the green dashed line at $\sim 10^{20}$ cm in the figure) is located at SNR RX J1713. The CRs are assumed to enter the cloud from the outer edge, while the UV emission originates from sources at the inner edge of the ISM ring.

## APPENDIX H: CLOUDY

This appendix presents the individual [C II] emissivity contributions from the modelled UV sources. Figs H1 and H2 show the emissivity from the O9.5 stars and the H II/star-forming regions separately for hydrogen densities of 1 and 100 $\mathrm{cm^{-3}}$, respectively.

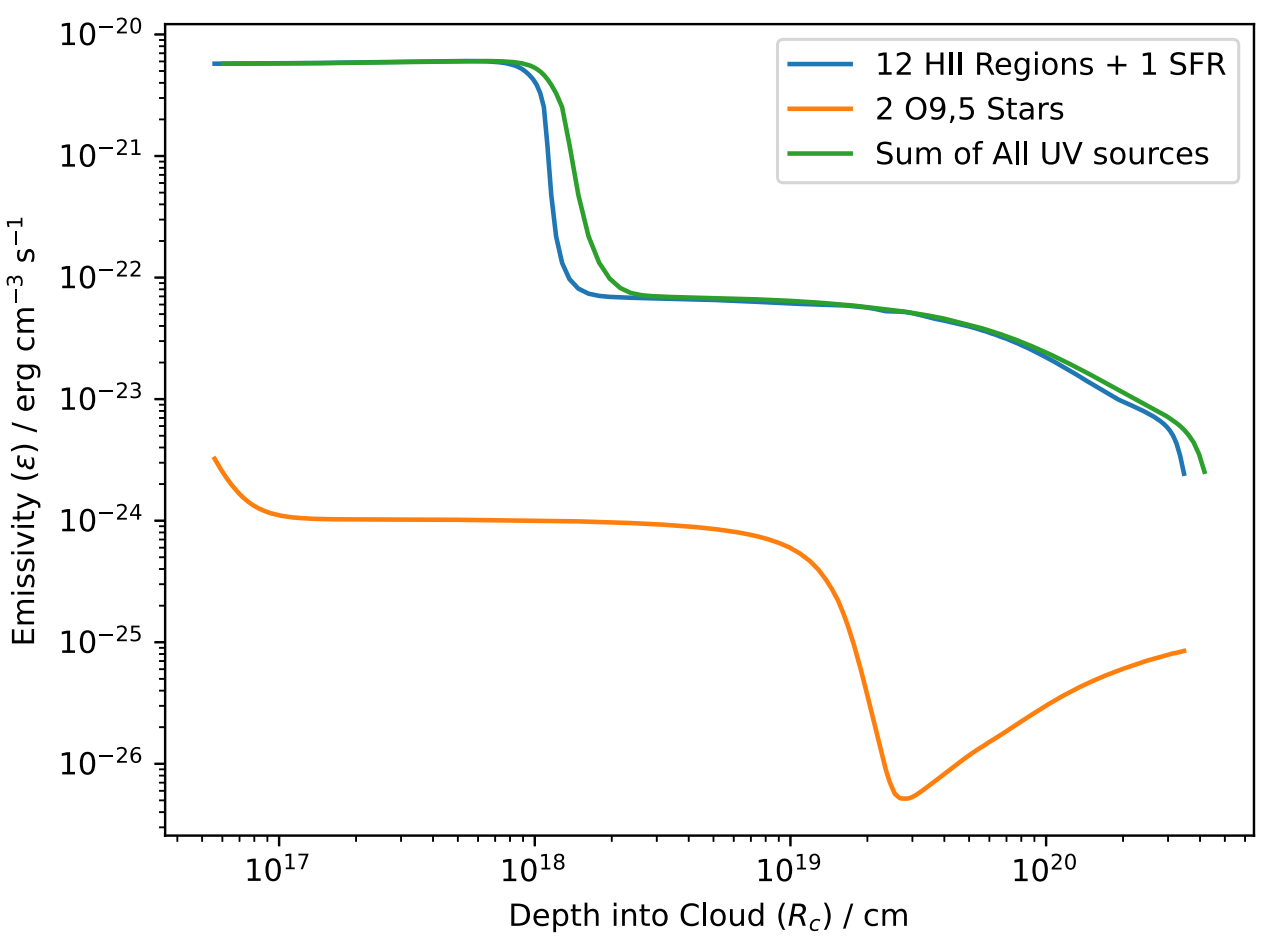

**Figure H2.** [C II] emission predicted by CLOUDY at SNR RX J1713, versus depth of ISM shell $R_c$ with $n_H = 100\,\mathrm{cm^{-3}}$. The blue line shows the [C II] emissivity from 2 O9.5 UV stars (100 pc away from the SNR), while the orange line shows the [C II] emissivity from the H II and star-forming regions (40 pc away). The total [C II] emissivity from all UV sources is shown by the green line. The *x*-axis shows the depth ($R_c$)into the shell of the cloud, where the outer edge of the cloud (shown by the green dashed line at $\sim 10^{20}$ cm in the figure) is located at SNR RX J1713. The CRs are assumed to enter the cloud from the outer edge, while the UV emission originates from sources at the inner edge of the ISM ring.

## APPENDIX I: UV CONTINUUM CONSTRAINTS TOWARDS RX J1713.7−3946

To constrain the ultraviolet (UV) radiation field towards RX J1713.7−3946, we derived an upper limit on the continuum flux density around 3000 Å from *GALEX* observations in a nearby pointing at $\sim 347.3^\circ$ Galactic longitude and $-0.5^\circ$ Galactic latitude (P. Morrissey et al. 2007).

$$F_{\lambda,\mathrm{GALEX}}(3000\,\text{Å}) = 2.096 \times 10^{-13}\ \mathrm{erg\,cm^{-2}\,s^{-1}\,\text{Å}^{-1}}. \qquad \text{(I1)}$$

The [C II] photoionization models were computed with CLOUDY, which provides the total emergent continuum luminosity $L_\lambda$ (integrated over the modelled region) as a function of wavelength in microns. The total continuum luminosity corresponds to the sum of transmitted, diffuse and reflected continuum, and is therefore the appropriate quantity to compare with the observed flux from *GALEX*.

At a wavelength closest to 0.3 μm, the continuum data yield an intrinsic (unextincted) luminosity of

$$L_{\mathrm{CLOUDY,intr}}(0.3\ \mu\mathrm{m}) = 1.110 \times 10^{38}\ \mathrm{erg\,s^{-1}\,\mu m^{-1}}. \qquad \text{(I2)}$$

Assuming a distance of 1 kpc to RX J1713.7−3946, the corresponding intrinsic flux at Earth is

$$F_{\mathrm{CLOUDY,ext}}(3000\,\text{Å}) = \frac{L_{\mathrm{CLOUDY,intr}} \times 10^{-0.4\times A_{3000}}}{4\pi d^2 \times 10^4} \simeq 1.8 \times 10^{-13}\ \mathrm{erg\,cm^{-2}\,s^{-1}\,\text{Å}^{-1}}, \qquad \text{(I3)}$$

where $A_{3000} \simeq 1.7 \times A_V$. To account for Galactic foreground extinction, we adopt the Milky Way extinction law of J. A. Cardelli, G. C. Clayton, J. S. Mathis (1989). The shape of the extinction curve of the Milky Way is given by $R_V = 3.1$. We consider a foreground visual extinction of $A_V \simeq 4$ mag along the line of sight to RX J1713.7−3946 (Y. Fukui et al. 2012).

The observed *GALEX* upper limit is shown in equation (I1). The ratio of the extincted model flux to this limit is

$$\frac{F_{\lambda,\mathrm{ext}}}{F_{\lambda,\mathrm{GALEX}}} \simeq \frac{1.8 \times 10^{-13}}{2.1 \times 10^{-13}} \approx 0.85. \qquad \text{(I4)}$$

Thus, for a distance of 1 kpc and a foreground extinction of $A_V \simeq 4$ mag, the UV continuum predicted by the CLOUDY model at 3000 Å lies slightly below, but close to, the *GALEX* upper limit, implying that the UV constraint from *GALEX* does not strongly exclude the adopted model parameters but does place them near the observational limit.